\documentclass[acmsmall,screen=true,anonymous=false,bookmarks=false, nonacm]{acmart}

\usepackage{blkarray}                                      
\usepackage[noend]{algpseudocode}
\usepackage{algorithm}

\usepackage{graphicx}
\usepackage{amsmath}
\usepackage{amsthm}
\usepackage{booktabs}
\usepackage{enumerate}
\usepackage{multirow}
\usepackage[subrefformat=parens,farskip=0pt,justification=centering]{subfig}

\usepackage{color}
\usepackage{comment}
\usepackage{soul}
\soulregister\cite7
\soulregister\ref7
\soulregister\pageref7
\usepackage{etoolbox}
\usepackage{url}
\usepackage{nth}
\usepackage{bm}
\usepackage{balance}
\usepackage{threeparttable}
\usepackage{xcolor,colortbl}
\usepackage{footnote}
\usepackage{listings}
\usepackage{setspace}
\usepackage{verbatim}
\hypersetup{
    colorlinks = true,
    citecolor  = CUHKpurple,
    linkcolor  = CUHKpurple,
    urlcolor   = CUHKpurple,
}
\usepackage{tikz}
\usetikzlibrary{patterns}
\usetikzlibrary{positioning,calc,fit,decorations.pathmorphing,shapes.geometric, shapes.gates.logic.US, calc}
\usetikzlibrary{arrows,arrows.meta,decorations.markings,shapes,shapes.arrows}
\usetikzlibrary{decorations,decorations.pathreplacing}
\usetikzlibrary{backgrounds}
\usepackage{pgfplots}
\usepackage{pgfplotstable}
\usepackage{scalefnt}
\pgfplotsset{compat=newest}
\usepackage{caption}
\usepackage{cleveref}
\Crefformat{figure}{Fig.~#2#1#3}
\Crefname{subfigure}{Fig.}{Figs.}
\Crefname{figure}{Fig.}{Figs.}
\Crefformat{table}{TABLE~#2#1#3}
\usepackage[figuresright]{rotating}
\usepackage{siunitx}
\usepackage{minted}
\usepackage{newfloat}
\usepackage{enumitem}

\definecolor{CUHKorange}{RGB}{244,106,18} 
\definecolor{CUHKblue}{RGB}{0,111,190}    
\definecolor{CUHKgreen}{RGB}{0,127,128}   
\definecolor{CUHKred}{RGB}{228,46,36}     
\definecolor{CUHKyellow}{RGB}{198,148,34} 
\definecolor{CUHKdark}{RGB}{114,44,114}   
\definecolor{CUHKmiddle}{RGB}{144,44,144} 
\definecolor{CUHKlight}{RGB}{167,44,167}
\definecolor{CUHKpurple}{RGB}{117,15,109}
\definecolor{CUHKgold}{RGB}{221,163,0}
\definecolor{CUHKribbon}{RGB}{244,223,176}
\definecolor{CUHKblack}{RGB}{34,24,21}

\renewcommand{\texttt}[1]{{\ttfamily\selectfont #1}}

\definecolor{MYred}{RGB}{255,207,206}
\definecolor{MYsnow}{RGB}{213,248,241}
\definecolor{MYgreen}{RGB}{212,239,194}
\definecolor{MYpurple}{RGB}{199,195,239}
\definecolor{MYyellow}{RGB}{230,209,139}
\definecolor{MYorange}{RGB}{235,161,146}
\definecolor{MYblue}{RGB}{176,235,231}
\definecolor{MYgrey}{RGB}{188,193,204}
\definecolor{vblue}{RGB}{49,49,255}
\definecolor{vorange}{RGB}{255,143,102}
\definecolor{vgreen}{RGB}{104,180,104}
\definecolor{darkgreen}{rgb}{0.0, 0.5, 0.0}
\definecolor{IRHw}{HTML}{0B5394}
\definecolor{IRSeq}{HTML}{2F6B2F}
\definecolor{IRPipe}{HTML}{B45F06}
\definecolor{IRComb}{HTML}{3D85C6}
\definecolor{IRType}{HTML}{666666}
\definecolor{IRAttr}{HTML}{C00000}
\definecolor{IRNum}{HTML}{C55A11}
\definecolor{IRSSA}{HTML}{444444}

\newcommand{\irpipe}[1]{\textcolor{IRPipe}{\texttt{#1}}}
\newcommand{\ircomb}[1]{\textcolor{IRComb}{\texttt{#1}}}
\newcommand{\irtype}[1]{\textcolor{IRType}{\texttt{#1}}}

\newcommand{\irnum}[1]{\textcolor{IRNum}{\texttt{#1}}}
\newcommand{\irssa}[1]{\textcolor{IRSSA}{\texttt{#1}}}

\newcommand{\minisection}[1]{\par\noindent\textbf{#1}.\enspace}

\usepackage{tcolorbox}
\tcbuselibrary{skins,breakable}
    {\endtcolorbox}
    {\endtcolorbox}

\crefname{mytheorem}{Theorem}{Theorems}
\crefname{mylemma}{Lemma}{Lemmas}
\crefname{myclaim}{Claim}{Claims}
\crefname{myproperty}{Property}{Properties}
\crefname{mycorollary}{Corollary}{Corollaries}

\algrenewcommand\textproc{\texttt}
\newcommand{\tabincell}[2]{
    \begin{tabular}{@{}#1@{}}#2\end{tabular}
}

\makeatletter
\let\OldStatex\Statex
\renewcommand{\Statex}[1][3]{%
  \setlength\@tempdima{\algorithmicindent}%
  \OldStatex\hskip\dimexpr#1\@tempdima\relax
}
\makeatother

\algblockdefx[Foreach]{Foreach}{EndForeach}[1]{\textbf{for each} #1 \textbf{do}}{\textbf{end for}}
\makeatletter
\ifthenelse{\equal{\ALG@noend}{t}}{\algtext*{EndForeach}}{}
\makeatother

\RequirePackage[normalem]{ulem}
\RequirePackage{color}\definecolor{RED}{rgb}{1,0,0}\definecolor{BLUE}{rgb}{0,0,1}

\makeatletter
\newcommand*\@lbracket{[}
\newcommand*\@rbracket{]}
\newcommand*\@colon{:}
\newcommand*\colorIndex{%
    \edef\@temp{\the\lst@token}%
    \ifx\@temp\@lbracket \color{black}%
    \else\ifx\@temp\@rbracket \color{black}%
    \else\ifx\@temp\@colon \color{black}%
    \else \color{vorange}%
    \fi\fi\fi
}
\makeatother

\lstdefinestyle{verilog-style}
{
    language=Verilog,
    basicstyle=\small\ttfamily,
    keywordstyle=\color{vblue},
    identifierstyle=\color{black},
    commentstyle=\color{vgreen},
    numbers=left,
    numberstyle=\tiny\color{black},
    numbersep=8pt,
    tabsize=8,
    moredelim=*[s][\colorIndex]{[}{]},
    literate=*{:}{:}1,
    frame=single,
    captionpos=b
}

\SetupFloatingEnvironment{listing}{name=Code}

\setminted{
    frame = lines,
    fontsize = \small,
    linenos,
    breaklines,
    xleftmargin = 0.015\textwidth,
    xrightmargin = 0.015\textwidth,
    baselinestretch = 0.8
}
\crefname{lstlisting}{code}{codes}
\Crefname{lstlisting}{Code}{Codes}
\newcommand*\circled[1]{\tikz[baseline=(char.base)]{%
            \node[shape=circle,color=blue,fill=blue!20,draw,inner sep=0.6pt] (char) {#1};}}
\graphicspath{{./figs/}{../}}

\begin{document}

\title{
    PipeRTL: Timing-Aware Pipeline Optimization at IR-Level for RTL Generation
}

\acmJournal{TACO}
\settopmatter{printfolios=true,authorsperrow=5}
\setcopyright{none}
\renewcommand\shortauthors{Yin et al.}

\author{Shuo Yin}
\affiliation{
  \department{Department of Computer Science and Engineering}
  \institution{The Chinese University of Hong Kong}
  \city{Hong Kong}
  \country{Hong Kong SAR}
}
\email{yinshuo991229@link.cuhk.edu.hk}

\author{Fangzhou Liu}
\affiliation{%
  \department{Department of Computer Science and Engineering}
  \institution{The Chinese University of Hong Kong}
  \city{Hong Kong}
  \country{Hong Kong SAR}
}
\email{fzliu23@cse.cuhk.edu.hk}

\author{Lancheng Zou}
\affiliation{%
  \department{Department of Computer Science and Engineering}
  \institution{The Chinese University of Hong Kong}
  \city{Hong Kong}
  \country{Hong Kong SAR}
}
\email{lczou23@cse.cuhk.edu.hk}

\author{Rongliang Fu}
\affiliation{%
  \department{Department of Computer Science and Engineering}
  \institution{The Chinese University of Hong Kong}
  \city{Hong Kong}
  \country{Hong Kong SAR}
}
\email{rlfu@cse.cuhk.edu.hk}

\author{Wenqian Zhao}
\affiliation{%
  \department{Department of Computer Science and Engineering}
  \institution{The Chinese University of Hong Kong}
  \city{Hong Kong}
  \country{Hong Kong SAR}
}
\email{wqzhao@cse.cuhk.edu.hk}

\author{Chen Bai}
\authornote{Corresponding authors.}
\affiliation{%
  \department{Department of Electronic and Computer Engineering}
  \institution{The Hong Kong University of Science and Technology}
  \city{Hong Kong}
  \country{Hong Kong SAR}
}
\email{eecbai@ust.hk}

\author{Tsung-Yi Ho}
\affiliation{%
  \department{Department of Computer Science and Engineering}
  \institution{The Chinese University of Hong Kong}
  \city{Hong Kong}
  \country{Hong Kong SAR}
}
\email{tyho@cse.cuhk.edu.hk}

\author{Yuan Xie}
\affiliation{%
  \department{Department of Electronic and Computer Engineering}
  \institution{The Hong Kong University of Science and Technology}
  \city{Hong Kong}
  \country{Hong Kong SAR}
}
\email{yuanxie@ust.hk}

\author{Bei Yu}
\authornotemark[1]
\affiliation{%
  \department{Department of Computer Science and Engineering}
  \institution{The Chinese University of Hong Kong}
  \city{Hong Kong}
  \country{Hong Kong SAR}
}
\email{byu@cse.cuhk.edu.hk}

\begin{abstract}

Modern hardware compilers increasingly rely on rich intermediate representations (IRs) to preserve optimization-relevant semantics before generating RTL code. 
However, one important optimization is still largely deferred to backend tools: pipeline optimization. In common RTL flows, registers are inserted by frontend heuristics or hardware designers and later adjusted by backend retiming after the design has been lowered to a much lower-level netlist representation.
At that point, much of the operator-level structure originally exposed by the compiler IR has already been weakened or lost, limiting opportunities for global, compiler-level pipeline optimization.

This paper presents \textit{PipeRTL}, an IR-level pipeline optimization framework for hardware compilers, instantiated in CIRCT. 
PipeRTL makes the legality of register relocation explicit in the IR, uses a learned timing predictor to approximate downstream delay behavior, and formulates timing-aware register relocation as a global min-cost flow problem under timing constraints. 
Evaluation on open-source designs under a commercial backend synthesis flow shows that PipeRTL improves downstream implementation quality on average, reducing critical-path delay, power, and area across the evaluated benchmarks, while also providing a stronger starting point for backend retiming.
These results indicate that exposing pipeline optimization as an explicit compiler pass can deliver backend-meaningful gains by improving the sequential structure presented to later stages and the resulting downstream implementation quality.

\end{abstract}

\ccsdesc[500]{Software and its engineering~Compilers}
\ccsdesc[300]{Hardware~Logic design}
\keywords{hardware compiler, RTL generation, pipeline optimization, timing-aware optimization, CIRCT}

\maketitle
\pagestyle{plain}

\section{Introduction}
\label{sec:introduction}

As modern hardware becomes increasingly complex, many hardware design languages (HDLs)~\cite{weng2025assassyn, xiao2024cement, chen2024allo, EDA-2014MICRO-PyMTL} and hardware intermediate representations (IRs)~\cite{lai2019heterocl, schuiki2020llhd, nigam2021compiler, Li:EECS-2016-9} have been developed to provide higher-level abstractions and preserve semantic information for optimization and transformation before RTL generation.
Yet one important optimization is still largely deferred to backend tools: pipeline optimization. Pipelining determines where registers are placed and therefore where the critical path can emerge. Beyond timing, pipeline structure also affects power and area through its impact on register usage and logic organization. Even when multiple RTL implementations are functionally equivalent, their register placements can expose very different optimization opportunities.

Existing RTL generation approaches, including high-level synthesis (HLS) flows~\cite{ferrandi2021,ye2022scalehls,cheng2024seer,nigam2021compiler,jo2020soff, canis2011legup} and agile hardware development frameworks~\cite{EDA-2012DAC-Chisel,Misc-SpinalHDL,bourgeat2020essence,mashtizadeh2007phdl}, usually determine pipeline structure without a downstream-aware optimization loop.
HLS tools typically insert registers based on software-level control flow synthesis, while agile hardware frameworks often rely on human intuition and experience to place registers in the RTL.
These frontend decisions are later revisited by backend pipeline optimization strategies such as retiming~\cite{leiserson1991retiming,shenoy2003efficient}, which operate after logic synthesis and technology mapping in a netlist representation, when much of the operator-level structure originally exposed by the compiler IR has already been weakened or lost.
As a result, backend retiming must search over a lower-level representation with less explicit semantic guidance, relying heavily on heuristics to navigate a difficult optimization space.


We argue that the compiler IR level provides an effective point for pipeline optimization, bridging the gap between frontend and backend.
High-level hardware IRs before lowering can still preserve optimization-relevant semantics that are difficult to recover later, including operation boundaries, stage relationships, and explicit fan-in and fan-out roles, providing richer guidance than a flattened netlist.
At the same time, the IR is close enough to the implementation that bitwidth-dependent cost differences, register types, and stage potentials can be reasoned about explicitly, making pipeline optimization a structured compiler problem rather than either an uninformed frontend decision or a low-level backend recovery step.

However, achieving effective IR-level pipeline optimization introduces several key challenges.
First, the IR level lacks detailed physical information about operators, making it difficult to estimate the timing effect of register relocation. Such information depends on the process design kit and physical implementation of the circuit, which are usually available only in backend flows.
Second, pipeline optimization must model how registers move across combinational 
logic, but current hardware IRs generally do not provide an explicit representation 
of the legality conditions for such transformations.

Motivated by these challenges, we present \textit{PipeRTL}, a compiler-guided pipeline optimization framework for hardware IRs, instantiated in Circuit IR Compilers and Tools (CIRCT)~\cite{Misc-CIRCT}.
PipeRTL models register relocation as an IR-level transformation over a semantically enriched representation. 
This representation makes pipeline legality explicit through new semantics for stage potentials, delay boundaries, and source/sink constraints. 
On top of this representation, PipeRTL uses a learned timing cost model to approximate downstream timing behavior and formulates timing-aware register relocation as a min-cost flow problem under timing constraints. 
PipeRTL therefore treats pipeline optimization as a general compiler pass with explicit representation, legality, cost modeling, and global optimization, rather than leaving it entirely to backend heuristics.

Our experiments on open-source processors and accelerators show that PipeRTL improves downstream implementation quality on average under a commercial backend synthesis flow.
These gains are meaningful in a backend-facing setting, where even single-digit improvements in power or area after timing-constrained synthesis are typically considered nontrivial.
In addition, PipeRTL provides a stronger starting point for later retiming and physical optimization in our evaluated backend flow, further improving design quality. More broadly, this perspective matters for hardware compilation as a whole.
Rather than viewing sequential optimization only as a backend recovery step, we show that it can also be exposed as a compiler optimization problem with explicit IR semantics, analyzable legality conditions, and a practical integration point with downstream implementation flows.
At the same time, PipeRTL does not aim to replace backend optimization or exact physical timing analysis. 
Its role is to improve the quality of the sequential structure presented to later stages, especially in designs whose pipeline organization is still malleable at the IR level.
The main contributions of this work are summarized as follows:
\begin{itemize}
     \item We formulate pipeline optimization as a compiler middle-end problem over hardware IR, rather than treating it solely as a backend retiming task after lowering.
     \item We design an IR representation that makes the legality of register relocation explicit, including stage potentials, delay boundaries, and source/sink constraints needed to preserve sequential behavior.
     \item We develop a practical optimization flow that couples learned timing estimation with min-cost-flow-based register relocation, enabling global pipeline restructuring under timing constraints.
     \item We implement the approach in CIRCT and evaluate it on open-source processor and accelerator designs, showing average downstream PPA benefits under a commercial backend flow, including additional gains when combined with backend retiming in our evaluated setting.
\end{itemize}

\section{Motivation \& Background}
\label{sec:background-and-motivation}

\begin{figure}[t]
    \centering
    \resizebox{\linewidth}{!}{
        \begin{minipage}[b]{0.323\linewidth}
            \centering
\begin{lstlisting}[
                style={verilog-style},
                basicstyle=\fontsize{5.5}{6.3}\selectfont\ttfamily,
                numberstyle=\fontsize{4}{4.6}\selectfont\color{black},
                numbersep=3pt,
                xleftmargin=1.2em,
                framexleftmargin=1.2em,
                caption=Example \uppercase\expandafter{\romannumeral1},
                label=lst:t0
            ]
module IIR(
  input        clock,
  input  [7:0] io_x,
  output [7:0] io_y
);

  reg [7:0] reg_0, reg_1, reg_2, 
            reg_3, reg_4;

  wire [7:0] wire_0 = reg_3 + reg_4;  
  wire [7:0] wire_1 = a * reg_0; 
  wire [7:0] wire_2 = b * reg_2; 

  assign io_y = io_x + wire_0; 

  always @(posedge clock) begin
    reg_0 <= io_x + wire_0;
    reg_1 <= io_x + wire_0;
    reg_2 <= reg_1;
    reg_3 <= wire_1;
    reg_4 <= wire_2;
  end
endmodule
\end{lstlisting}
        \end{minipage}
        \hspace{0.0065\linewidth}
        \begin{minipage}[b]{0.323\linewidth}
            \centering
\begin{lstlisting}[
                style={verilog-style},
                basicstyle=\fontsize{5.5}{6.3}\selectfont\ttfamily,
                numberstyle=\fontsize{4}{4.6}\selectfont\color{black},
                numbersep=3pt,
                xleftmargin=1.2em,
                framexleftmargin=1.2em,
                caption=Example \uppercase\expandafter{\romannumeral2},
                label=lst:t1
            ]
module IIR(
  input        clock,
  input  [7:0] io_x,
  output [7:0] io_y
);

  reg [7:0] reg_0, reg_1, reg_2, 
            reg_3;

  wire [7:0] wire_0 = reg_2 + reg_3;  
  wire [7:0] wire_1 = a * reg_0; 
  wire [7:0] wire_2 = b * reg_1; 

  assign io_y = io_x + wire_0; 

  always @(posedge clock) begin
    reg_0 <= io_x + wire_0;
    reg_1 <= reg_0;
    reg_2 <= wire_1;
    reg_3 <= wire_2;
  end
  
endmodule
\end{lstlisting}
        \end{minipage}
        \hspace{0.0065\linewidth}
        \begin{minipage}[b]{0.323\linewidth}
            \centering
\begin{lstlisting}[
                style={verilog-style},
                basicstyle=\fontsize{5.5}{6.3}\selectfont\ttfamily,
                numberstyle=\fontsize{4}{4.6}\selectfont\color{black},
                numbersep=3pt,
                xleftmargin=1.2em,
                framexleftmargin=1.2em,
                caption=Example \uppercase\expandafter{\romannumeral3},
                label=lst:t2
            ]
module IIR(
  input        clock,
  input  [7:0] io_x,
  output [7:0] io_y
);

  reg [7:0] reg_0, reg_1, reg_2; 


  wire [7:0] wire_0 = a * reg_0;
  wire [7:0] wire_1 = b * reg_1; 


  assign io_y = io_x + reg_2; 

  always @(posedge clock) begin
    reg_0 <= io_x + reg_2;
    reg_1 <= reg_0;
    reg_2 <= wire_0 + wire_1;
  end

  
endmodule
\end{lstlisting}
        \end{minipage}
    }
    \caption{Three equivalent implementations of an IIR filter.}
    \label{fig:example}
\end{figure}

\subsection{Pipeline Synthesis Methodologies}
\label{subsec:pipeline-synthesis}
Pipeline synthesis determines where registers are introduced or relocated so that a circuit can meet throughput and timing goals.
Existing methodologies largely fall into three groups.
First, in agile hardware design languages~\cite{EDA-2012DAC-Chisel,weng2025assassyn,mashtizadeh2007phdl,bourgeat2020essence}, pipeline structure is often chosen by designers and encoded directly in the generated RTL.
Event-based HDLs~\cite{xiao2024cement, nigam2026parameterized} make this structure more explicit through specialized semantics, but placement decisions are still largely driven by designer intent rather than compiler optimization.
Second, HLS frameworks~\cite{ye2022scalehls,lai2019heterocl, elakhras2025elasticmiter} derive pipeline structure from scheduling and control-flow synthesis.
Prior work has further explored basic-block pipelining and SDC-based formulations~\cite{josipovic2021buffer, ye2024subgraph}.
These approaches are effective at the scheduling level, but their optimization units are typically coarse and do not directly expose fine-grained, legality-preserving register relocation opportunities over an RTL-oriented IR.
Third, synthesis-oriented approaches such as e-graph-based exploration~\cite{coward2024rover, cheng2024seer} broaden the search space for structural rewrites and pipelining choices.
However, they do not provide an explicit compiler representation of sequential legality together with a timing-aware objective for global register movement.

The gap, therefore, is not simply a lack of pipelining support, but a lack of a compiler-stage methodology that treats pipeline optimization itself as an IR-level transformation problem.
PipeRTL targets this gap by making sequential legality explicit in the IR, reasoning about downstream timing impact during optimization, and performing global register relocation before lowering to less structured representations.

\minisection{Motivation Example}
Using an example, we illustrate how register placement affects RTL performance.
\Cref{eq:iir} shows the formulation of a one-dimensional infinite impulse response (IIR) filter using 8-bit integer data.
\begin{equation}
	\label{eq:iir}
	\begin{aligned}
		w(n) &= a \cdot y(n-1) + b \cdot y(n - 2), \\
		y(n) &= w(n - 1) + x(n).
	\end{aligned}
\end{equation} 
Accordingly, \Cref{lst:t0}, \Cref{lst:t1}, and \Cref{lst:t2} show three Verilog implementations that differ only in register placement.
Although the three implementations satisfy \Cref{eq:iir}, their register usage and critical paths differ, as detailed in \Cref{tbl:iir}.
In Example \uppercase\expandafter{\romannumeral1}, the implementation follows \Cref{eq:iir} directly and uses $5$ registers (line $7$).
The critical path is defined by lines $10$ and $17$, which contain two cascaded $8$-bit adders; for this motivating example, we assume that two cascaded 8-bit adders have greater latency than one 8-bit multiplier.
In Example \uppercase\expandafter{\romannumeral2}, combining one delay of $y(n - 1)$ and $y(n - 2)$ in the definition of $w(n)$ reduces the register count to $4$. The critical path remains the same as in Example \uppercase\expandafter{\romannumeral1}: two cascaded $8$-bit adders.
In contrast, Example \uppercase\expandafter{\romannumeral3} moves the register in Example \uppercase\expandafter{\romannumeral2} from the fan-in side of the $8$-bit adder (line $10$) to its fan-out side, further reducing the register count to $3$. However, the critical path then contains an $8$-bit multiplier and an $8$-bit adder, which is worse than in Examples \uppercase\expandafter{\romannumeral1} and \uppercase\expandafter{\romannumeral2}.
These three examples show that register relocation changes both resource usage and timing behavior even when the RTL remains functionally equivalent.
Some relocations reduce the number of registers without affecting the critical path, while others improve one metric at the expense of another.
This illustrates the compiler problem addressed by PipeRTL: how to search this space of legal sequential reorganizations systematically, while preserving functionality and avoiding downstream timing degradation.

\begin{table}[tb!]
    \centering
    \caption{Differences among the three IIR implementations, where \#Reg denotes register count.}
    \label{tbl:iir}
    \resizebox{0.38\linewidth}{!} {
        \begin{threeparttable} {
            \begin{tabular}{|c|c|c|}
                \hline
                \textbf{Code}      & \textbf{\#Reg} 		& \textbf{Critical Path} \\ \hline \hline
                Listing 1 & 5            &  \textcolor{green}{$+$\texttt{(8b)} $\to$ $+$\texttt{(8b)}}     \\
                Listing 2 & 4            &    \textcolor{green}{$+$\texttt{(8b)} $\to$ $+$\texttt{(8b)}}     \\
                Listing 3 & \textcolor{green}{3}            &    $\times$\texttt{(8b)} $\to$ $+$\texttt{(8b)}      \\ \hline
            \end{tabular}
            \begin{tablenotes}
                \small
                \item[1] $b$ bitwidth; $+$ adder; \\ $\times$ multiplier; $\to$ wire.
            \end{tablenotes}
        }
        \end{threeparttable} 
    }
\end{table}

\minisection{Circuit IR Compilers and Tools}
Circuit IR Compilers and Tools (CIRCT) \cite{Misc-CIRCT} is a comprehensive compiler stack for hardware design language implementation. Built on the MLIR/LLVM infrastructure, it supports applications such as high-level synthesis~\cite{nigam2021compiler}, Verilog/SystemVerilog generation~\cite{weng2025assassyn, xiao2024cement}, and simulation~\cite{EDA-2023MICRO-Khronos,schuiki2020llhd}.
CIRCT supports core dialects for RTL representation, which include:
\begin{itemize}
	\item \texttt{hw} dialect offers function-like semantics to represent module information. 
	For instance, \textit{hw.module} captures module definitions, while \textit{hw.instance} represents module instances.
	\item \texttt{comb} dialect represents combinational components in RTL.
	For example, the \textit{comb.add} operation can have a variadic number of operands and one result to represent an adder.
	\item \texttt{seq} dialect represents sequential logic. 
	\textit{seq.firreg} models a register lowered from FIRRTL \cite{Li:EECS-2016-9}, containing the piped value and reset as inputs. 
	\item \texttt{sv} dialect represents SystemVerilog semantics for RTL generation.
	For example, \textit{sv.always} represents an always block, typically used to define sequential logic. 
\end{itemize}
The core dialects of CIRCT serve as a middleware layer for representing RTL with multiple levels of abstraction, supporting flexible transformations and conversions.
This infrastructure offers a natural optimization point for PipeRTL before backend lowering. Leveraging CIRCT's structural preservation, PipeRTL introduces pipeline-specific semantics and analyses for register relocation.

\section{Overview of PipeRTL}
\label{sec:overview-of-PipeRTL}

\begin{figure*}
    \centering
    \includegraphics[width=.96\linewidth]{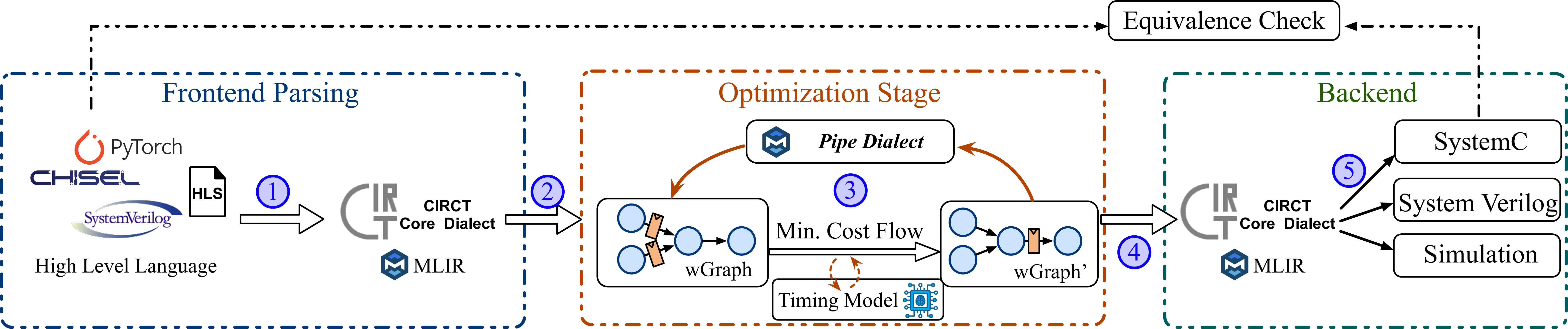}
    \caption{Overall framework of PipeRTL.}
    \label{fig:PipeRTL_framework}
\end{figure*}

\Cref{fig:PipeRTL_framework} shows the overall framework of PipeRTL, with the transformation steps \textcolor{blue}{highlighted}. 
PipeRTL transforms the circuit IR into a graph representation through a series of passes, performs pipeline optimization on the graph, and then lowers the optimized graph back into core dialects for downstream processing.

\begin{enumerate}[label=\protect\circled{\arabic*}]
 \item PipeRTL reuses CIRCT frontend infrastructure to translate input designs into CIRCT core dialects, so pipeline optimization can operate on a common IR regardless of the source language.
 \item PipeRTL converts the circuit IRs into the \texttt{Pipe} dialect, which models RTL as a weighted graph that captures stage boundaries.
 \item The \texttt{Pipe} dialect provides explicit semantics for pipeline stage updates, with optimization guided by a learned timing predictor and formulated as a min-cost flow problem under timing constraints.
 \item PipeRTL lowers the optimized \texttt{Pipe} IR back to standard CIRCT core dialects, yielding optimized pipeline structures.
 \item The optimized IR can then be processed through standard CIRCT passes for multiple backend targets, including RTL generation, simulation, and formal verification.
\end{enumerate}

\section{PipeRTL Framework Design}
\label{sec:PipeRTL-design}

This section presents the methodology of PipeRTL as an IR-level pipeline optimization framework for hardware compilers.

\begin{figure*}
    \centering
    \includegraphics[width=\linewidth]{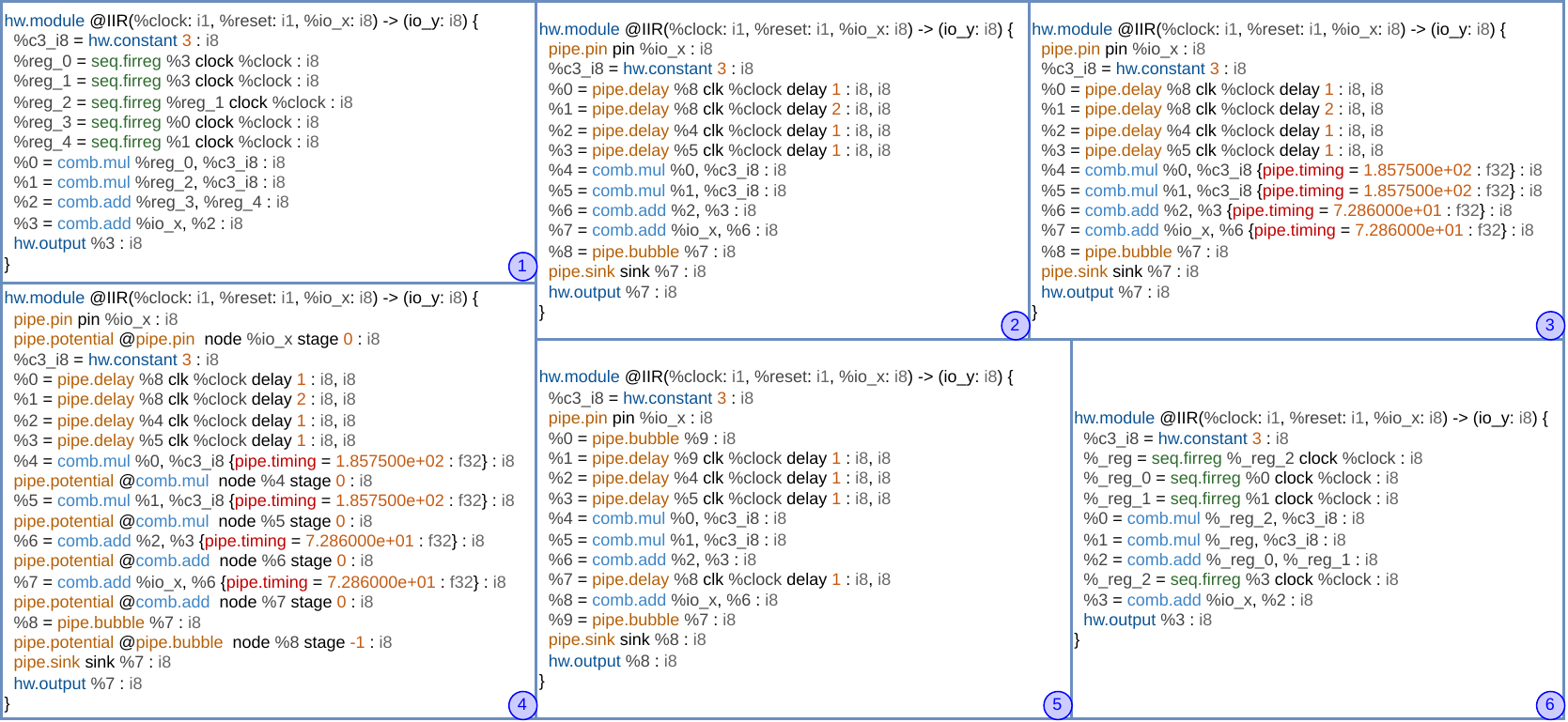}
    \caption{IR transformation flow of PipeRTL. Blue circles highlight the steps introduced by PipeRTL. The initial and optimized designs correspond to \Cref{lst:t0} and \Cref{lst:t1}, respectively.}
    \label{fig:ir}
\end{figure*}

\subsection{IR Representation in the \texttt{Pipe} Dialect}
\label{sec:pipe-dialect}

PipeRTL introduces a dedicated MLIR dialect, \texttt{Pipe}, to represent sequential structure in a form explicit enough for legality checking and global relocation. The dialect is integrated into CIRCT and works alongside existing CIRCT core dialects. \Cref{fig:ir} shows the transformation flow on the IIR example, where \Cref{fig:ir}(\circled{1}) is the initial CIRCT IR corresponding to \Cref{lst:t0} (coefficients $a$ and $b$ are initialized to 3).

This representation choice is important because the standard lowered forms used later in the flow no longer preserve enough structured information to express pipeline relocation directly as a compiler transformation. By introducing \texttt{Pipe}, PipeRTL keeps sequential structure explicit at the stage where operator boundaries, dataflow relationships, and boundary conditions are still naturally available to the compiler.

The central abstraction is a weighted directed graph $G(V, E)$ over the IR, which we call \textit{wGraph}. Nodes represent combinational operations or boundary nodes, and edges represent register-carrying or wire-like dependencies between them. \Cref{tbl:wgraph} summarizes how RTL constructs are mapped to wGraph components and corresponding \texttt{Pipe} operations.

\begin{table}[tb]
  \centering
  \caption{Mapping from RTL constructs to wGraph components and their corresponding \texttt{Pipe} operations.}
  \label{tbl:wgraph}
  \resizebox{0.6\linewidth}{!}
  {
    \begin{threeparttable}
        {
          \begin{tabular}{| c | c | c |}
            \hline
            \textbf{RTL Op} & \textbf{wGraph} & \textbf{Primitive} \\
            \hline \hline
            \tabincell{c}{
            \begin{minipage}[b]{0.25\columnwidth}
                \centering
                \raisebox{-.5\height}{\includegraphics[width=1.033\linewidth]{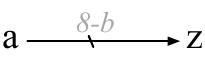}}
            \end{minipage}
            \\
            wire
         }
            & 
            \tabincell{c}{
            \begin{minipage}[b]{0.3\columnwidth}
                \centering
                
                \raisebox{-.5\height}{\includegraphics[width=0.85\linewidth]{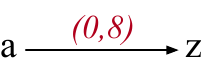}}
                 
            \end{minipage}
            \\
            $w(e_{a, z})$=0
            \\
            $\beta(e_{a, z})$=8
            }
            & 
            Use-define chain
            \\
            \hline
            \tabincell{c}{
            \begin{minipage}[b]{0.31\columnwidth}
                \centering
                \raisebox{-.5\height}{\includegraphics[width=1\linewidth]{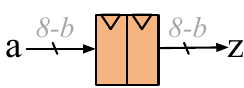}}
            \end{minipage}
            \\
            register
         }
            & 
            \tabincell{c}{
            \begin{minipage}[b]{0.25\columnwidth}
                \centering
          
                \raisebox{-.5\height}{\includegraphics[width=1.025\linewidth]{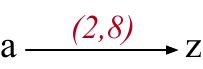}}
            \end{minipage}
                     \\
            $w(e_{a, z})$=2
            \\
            $\beta(e_{a, z})$=8
            }
            &
            \irssa{\%z} = \irpipe{pipe.delay} \irssa{\%a} clk \irssa{\%clk} delay \irnum{2} : \irtype{i8}
            \\
            \hline
            \tabincell{c}{
            \begin{minipage}[b]{0.25\columnwidth}
                \centering
          
                \raisebox{-.5\height}{\includegraphics[width=1.17\linewidth]{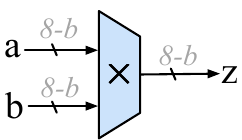}}
            \end{minipage}
            \\
            Comb logic
         }
            & 
            \tabincell{c}{
            \begin{minipage}[b]{0.25\columnwidth}
                \centering
          \resizebox{!}{.365\linewidth}{
                \raisebox{-.5\height}{\includegraphics[width=\linewidth]{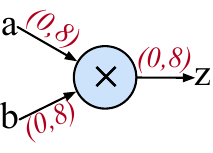}}}
            \end{minipage}
            \\
            Comb Node
        }
            & 
            \irssa{\%z} = \ircomb{comb.mul} \irssa{\%a}, \irssa{\%b} : \irtype{i8}
            \\
            \hline
            \tabincell{c}{
            \begin{minipage}[b]{0.25\columnwidth}
                \centering
                \raisebox{-.5\height}{\includegraphics[width=1.1\linewidth]{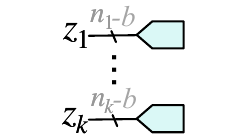}}
            \end{minipage}
            \\
            Output Port
        }
            & 
            \tabincell{c}{
            \begin{minipage}[b]{0.25\columnwidth}
                \centering
                
                \raisebox{-.5\height}{\includegraphics[width=0.68\linewidth]{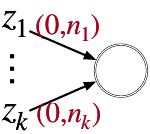}}
            \end{minipage}
            \\
            Sink Node
        }
            & 
            \irpipe{pipe.sink} sink \irssa{\%$z_1$} $\hdots$ \irssa{\%$z_k$} : \irtype{i3} $\hdots$ \irtype{i8}
            \\
            \hline
            \tabincell{c}{
            \begin{minipage}[b]{0.25\columnwidth}
                \centering
                \raisebox{-.5\height}{\includegraphics[width=1.47\linewidth]{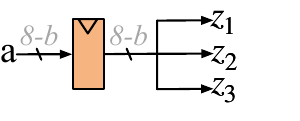}}
            \end{minipage}
            \\
            Broadcasting
        }
            & 
            \tabincell{c}{
            \begin{minipage}[b]{0.25\columnwidth}
                \centering
                \raisebox{-.5\height}{\includegraphics[width=1.18\linewidth]{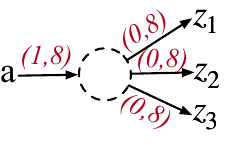}}
            \end{minipage}
            \\
            Bubble Node
        }
            & 
            \irssa{\%z} = \irpipe{pipe.bubble} \irssa{\%a} : \irtype{i8}
            \\
            \hline
            \tabincell{c}{
            \begin{minipage}[b]{0.25\columnwidth}
                \centering
                \raisebox{-.5\height}{\includegraphics[width=.9\linewidth]{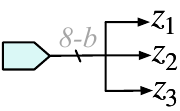}}
            \end{minipage}
             \\
            Input Port
         }
            & 
            \tabincell{c}{
            \begin{minipage}[b]{0.25\columnwidth}
                \centering
                \raisebox{-.5\height}{\includegraphics[width=0.775\linewidth]{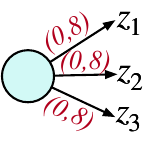}}
            \end{minipage}
            \\
            Pin Node
        }
            & 
            \irssa{\%z} = \irpipe{pipe.pin} pin \irssa{\%a} : \irtype{i8}
            \\
            \hline
          \end{tabular}
        }
    \end{threeparttable}
  }
\end{table}

In wGraph, each node $v$ represents either a combinational operation or a boundary node, and each edge $e_{u, v}$ captures the dependency between the fan-out of $u$ and the fan-in of $v$. Every edge carries two weights. The first, $w(e_{u, v})$, is the number of registers on that dependency and is encoded in the \texttt{Pipe} dialect by the delay of \texttt{pipe.delay}. If $w(e_{u, v})=0$, the edge is a wire-like connection with no intervening register. The second, $\beta(e_{u, v})$, denotes data capacity and is aligned with the IR type system.

Together, these two edge attributes let PipeRTL reason about both legality and cost in a single representation. The register-count weight captures where sequential boundaries can legally move, while the capacity weight links those moves to the register-resource objective used later in the optimization.

To preserve end-to-end behavior during relocation, PipeRTL introduces \texttt{pipe.pin} and \texttt{pipe.sink} to mark the boundaries of the circuit. These boundary operations make it explicit which paths must preserve their overall cycle latency and provide anchors for legality constraints in the optimization.

PipeRTL also introduces \texttt{pipe.bubble} to represent zero-delay broadcast structure. A bubble node has one fan-in and multiple fan-outs. Without it, a single register feeding multiple users would appear as several independent non-zero-weight edges and would therefore double count register capacity. The bubble abstraction preserves the sequential meaning of fan-out while keeping the optimization problem well formed.

\begin{figure}[htb]
    \centering
    \includegraphics[width=.6\linewidth]{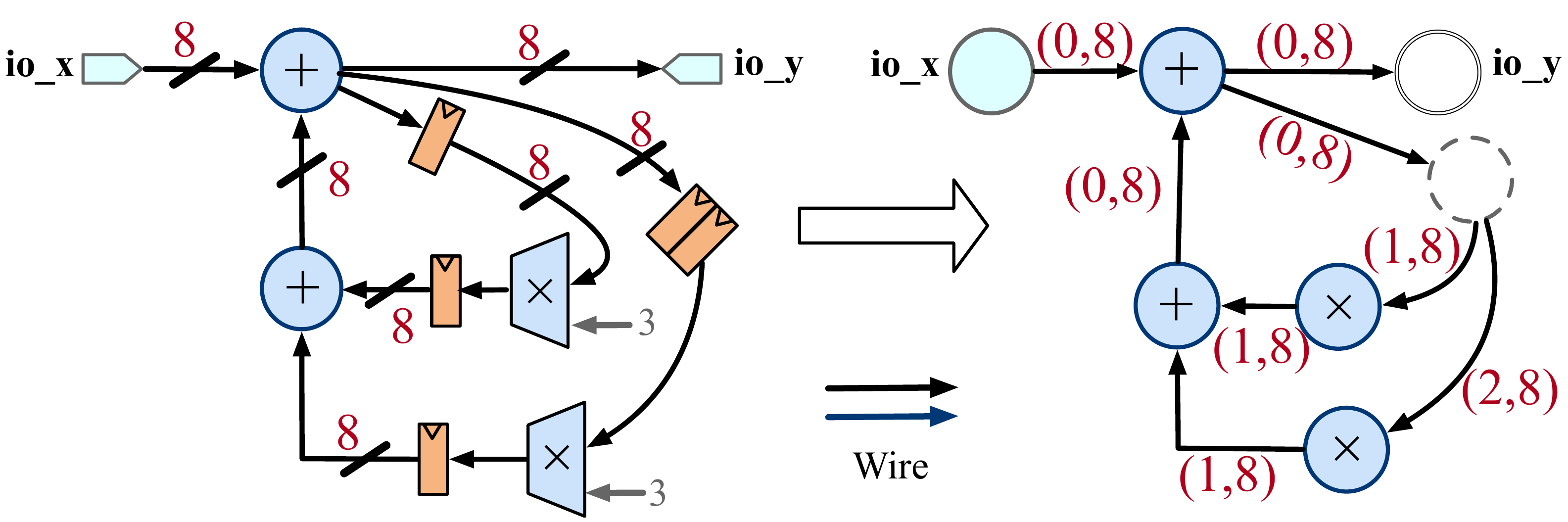}
    \caption{Example IIR design and the corresponding wGraph representation.}
    \label{fig:wgraph_example}
\end{figure}

From \Cref{fig:ir}(\circled{1}) to \Cref{fig:ir}(\circled{2}), PipeRTL rewrites registers into \texttt{pipe.delay}, inserts \texttt{pipe.bubble} where broadcast structure must be preserved, and marks \texttt{io\_x} and \texttt{io\_y} with \texttt{pipe.pin} and \texttt{pipe.sink}. \Cref{fig:wgraph_example} shows the resulting wGraph for the IIR example and illustrates how the representation exposes the sequential structure needed for legality-aware pipeline optimization.

External modules, such as IP blocks or SRAM macros, are treated as black boxes. Their outputs are modeled as \texttt{pipe.pin} operations and their inputs as \texttt{pipe.sink} operations, allowing PipeRTL to optimize surrounding logic without assuming visibility into the internal implementation of the black-box module.

This treatment is deliberately conservative. It avoids attributing unsupported relocation freedom to module internals while still exposing enough structure to optimize the logic around those black-box boundaries.

\subsection{IR-Level Pipeline Transformation}
\label{sec:ir-transformation}

Following \Cref{fig:ir}, PipeRTL performs pipeline optimization as an IR transformation pass over the \texttt{Pipe} dialect. The method has four components: timing prediction, legality-preserving combinational propagation, timing-aware global optimization, and lowering of the optimized sequential structure back to standard CIRCT dialects.

\subsubsection{Timing Prediction for Combinational Operations}
\label{subsubsec:xgboost-timing-predictor}

\begin{figure}[tb!]
    \centering
    \includegraphics[width=.7\linewidth]{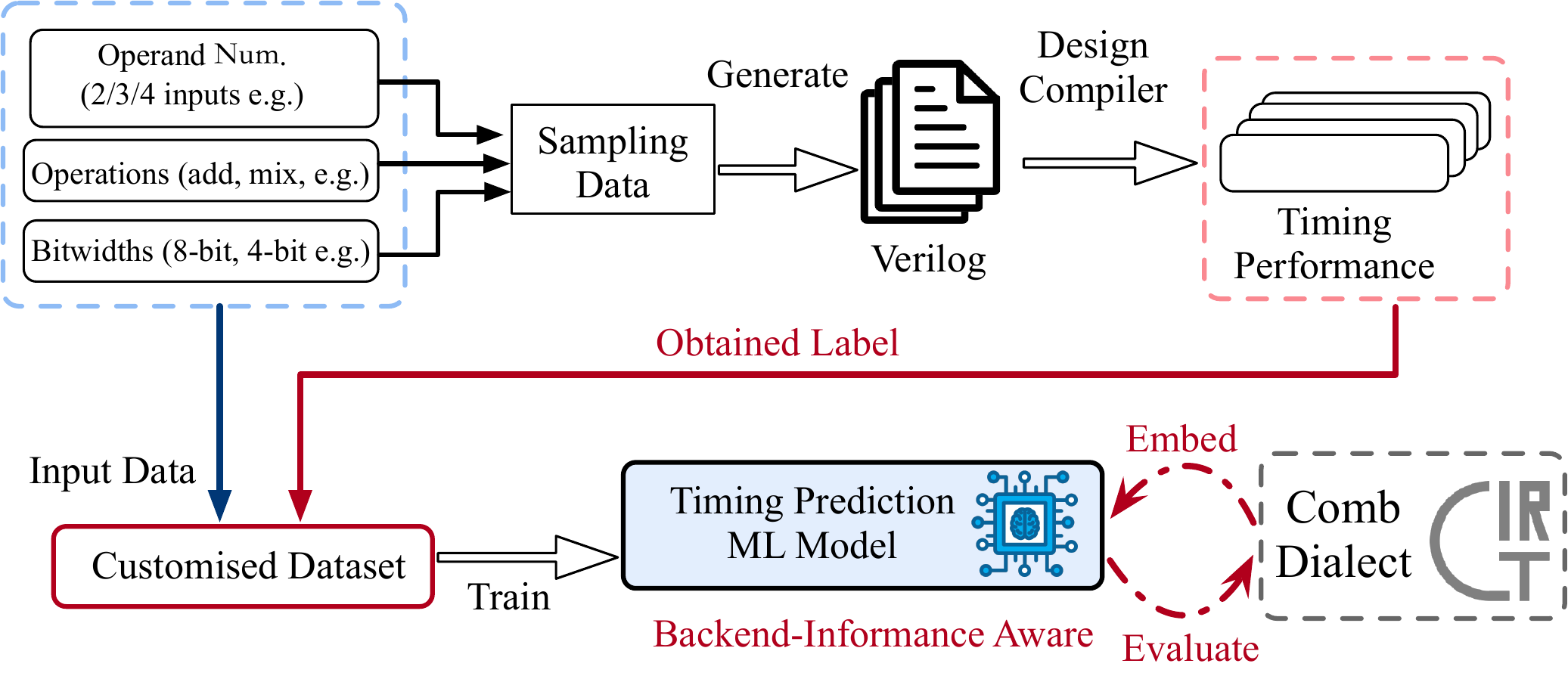}
    \caption{Training and inference flow of the XGBoost-based timing predictor.}
    \label{fig:timing_model}
\end{figure}

IR-level pipeline optimization requires a timing proxy for each combinational operation. A fixed delay lookup is insufficient because detailed gate-level information is unavailable at this stage and because the eventual implementation of an operation depends on factors such as operand count and bitwidth. For example, operations such as \texttt{comb.mul} and \texttt{comb.and} admit multiple implementations with different delay characteristics.

PipeRTL therefore uses a learned timing predictor to estimate a proxy delay $\delta(v)$ for each combinational node. We adopt XGBoost, which is widely used in machine-learning compiler cost models~\cite{chen2018tvm, chen2018learning, zheng2020ansor} due to its efficiency and interpretability.
As shown in \Cref{fig:timing_model}, we generate a Verilog training set with diverse combinational configurations, extract feature vectors from CIRCT \texttt{Comb} operations as $[\text{type},\ \text{number of operands},\ \text{bitwidth}]$, and use post-synthesis delay values from a commercial tool as labels. The predictor is trained as a regression model using mean squared error loss.

The goal of this predictor is not to reproduce full backend timing analysis inside the compiler. Instead, it provides a stable ranking signal that lets PipeRTL distinguish timing-critical from timing-insensitive combinational regions early enough to influence register placement decisions.

From \Cref{fig:ir}(\circled{2}) to \Cref{fig:ir}(\circled{3}), PipeRTL annotates each combinational operation with a \texttt{pipe.timing} attribute derived from this predictor. It then approximates the delay of a combinational path by summing the predicted delays of the nodes along that path. At this stage, approximate path delays are sufficient to guide pipeline restructuring; exact wire delay and physical implementation effects are intentionally deferred to later backend stages.

\begin{table*}[tbh!]
    \small
    \centering
    \caption{Two legal forms of combinational propagation represented in RTL and wGraph.}
    \label{tab:prop}
    \resizebox{\linewidth}{!}{
        \begin{threeparttable}
            \renewcommand{\arraystretch}{1.10} {
                \begin{tabular}{|c|c|c|c|c|c|}
                    \hline
                    \multicolumn{2}{|c|}{\textbf{Before Propagation}} & \multicolumn{2}{c|}{\textbf{After Propagation}} & \multirow{2}{*}{\textbf{$\Delta s(v)$}} & \multirow{2}{*}{\textbf{Primitive}} \\ \cline{1-4}
                    \multicolumn{1}{|c|}{RTL} & wGraph  & \multicolumn{1}{c|}{RTL} & wGraph & &\\
                    \hline \hline
                    \begin{minipage}[b]{0.28\columnwidth}
                        \centering
                        \raisebox{-.5\height}{\includegraphics[width=\linewidth]{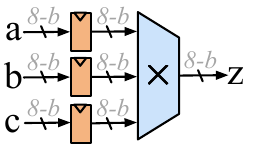}}
                    \end{minipage}
                    & 
                   \begin{minipage}[b]{0.24\columnwidth}
                        \centering
                        \raisebox{-.5\height}{\includegraphics[width=\linewidth]{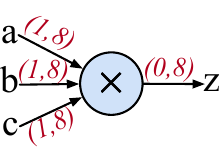}}
                    \end{minipage}
                   &
                    \begin{minipage}[b]{0.28\columnwidth}
                        \centering
                        \raisebox{-.5\height}{\includegraphics[width=\linewidth]{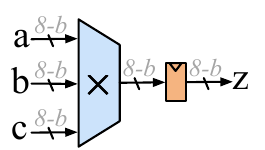}}
                    \end{minipage}
                    & 
                   \begin{minipage}[b]{0.24\columnwidth}
                        \centering
                        \raisebox{-.5\height}{\includegraphics[width=\linewidth]{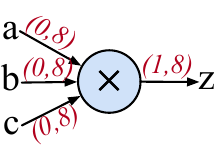}}
                    \end{minipage}
                    & \irnum{-1} & \irpipe{pipe.potential} \texttt{@mul node} \irssa{\%z} \texttt{stage} \irnum{-1} \texttt{:} \irtype{i8}
                    \\
                    \hline
                    \begin{minipage}[b]{0.28\columnwidth}
                        \centering
                        \raisebox{-.5\height}{\includegraphics[width=\linewidth]{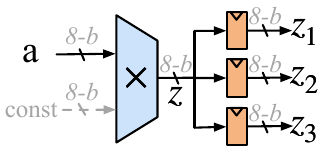}}
                    \end{minipage}
                    & 
                   \begin{minipage}[b]{0.24\columnwidth}
                        \centering
                        \raisebox{-.5\height}{\includegraphics[width=\linewidth]{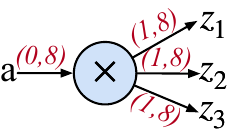}}
                    \end{minipage}
                   &
                    \begin{minipage}[b]{0.28\columnwidth}
                        \centering
                        \raisebox{-.5\height}{\includegraphics[width=\linewidth]{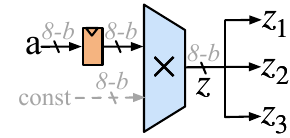}}
                    \end{minipage}
                    & 
                   \begin{minipage}[b]{0.24\columnwidth}
                        \centering
                        \raisebox{-.5\height}{\includegraphics[width=\linewidth]{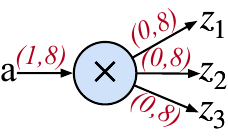}}
                    \end{minipage}
                    & \irnum{+1} & \irpipe{pipe.potential} \texttt{@mul node} \irssa{\%z} \texttt{stage} \irnum{1} \texttt{:} \irtype{i8}
                    \\
                    \hline
                \end{tabular}
            }
        \end{threeparttable}
    }
\end{table*}

\subsubsection{Combinational Propagation}

Given timing annotations, PipeRTL changes the sequential structure through \textit{combinational propagation}. Each combinational node $v$ is assigned a stage-update variable $\Delta s(v)$, which denotes how far the node is moved across surrounding registers. Moving a node forward by one stage is equivalent to pulling one register across the node from each fan-out edge to its fan-in edges; moving it backward performs the reverse transformation. \Cref{tab:prop} illustrates these two legal propagation directions.

This operation preserves functional behavior because it only changes where registers are placed relative to a combinational node, not the combinational computation itself. Pin and sink nodes are treated as circuit boundaries and therefore satisfy $\Delta s(v)=0$. For every path from a pin node to a sink node, the total cycle count remains unchanged after propagation. This gives PipeRTL a legality-preserving transformation primitive that can be reasoned about directly in the compiler middle-end.

This property is what makes combinational propagation suitable as the basic compiler transformation in PipeRTL. Rather than mutating the circuit through ad hoc rewrites, the pass manipulates a representation whose stage updates have an explicit interpretation and can therefore be checked systematically against boundary and timing constraints.

In the \texttt{Pipe} dialect, PipeRTL records the stage update of each node using the \texttt{pipe.potential} operation. This IR primitive serves as the carrier of pipeline relocation decisions during optimization and is later lowered into updated delay or bubble operations.

\subsubsection{Pipeline Optimization Formulation}
\label{subsec:pipeline-optimization-framework}

PipeRTL optimizes the pipeline by changing stage assignments of combinational operations while minimizing total register data capacity. Let $u, v \in V$ be two connected nodes in wGraph. The updated number of registers on edge $e_{u, v}$ after combinational propagation is
\begin{equation}
\label{eq:weight}
w'(e_{u, v}) = w(e_{u, v}) + \Delta s(v) - \Delta s(u).
\end{equation}
For legality, $w'(e_{u, v})$ must remain non-negative on every edge. The optimizer therefore searches for stage updates $\Delta s(v)$ that reduce weighted register capacity while preserving legal edge weights.

\minisection{Target Delay Searching}
Inspired by backend retiming~\cite{dc-retiming, leiserson1991retiming, shenoy2003efficient} that lets timing serve as a constraint instead of an objective, PipeRTL does not optimize register count independently of timing. Instead, it searches for a feasible target delay $T$ and only accepts stage updates whose induced path delays remain within that target.

PipeRTL performs this search as an outer feasibility loop over the wGraph. For a candidate target delay, it propagates delay estimates across zero-register paths, detects violations of the target, and incrementally adjusts stage updates to break overly long combinational regions. A candidate target is feasible only if the resulting stage updates satisfy both timing and legality, that is, all updated edge weights remain non-negative.
PipeRTL then uses binary search over the target delay to identify the tightest feasible timing target below the original critical path estimate.

This search is not intended to replace detailed backend timing analysis. Its role is to construct a timing-aware constraint set for IR-level pipeline relocation, so that the global optimizer operates over legal transformations that are also consistent with an estimated downstream timing objective.

In this sense, the target-delay search serves as the interface between local timing estimation and global register relocation. It converts predicted path delays into a compact set of admissible timing constraints that can be enforced in the subsequent optimization problem.

\minisection{Objective Formulation}
For any pair of nodes $u$ and $v$ in the wGraph, we define the minimal-weight path $u \rightsquigarrow v$ as the path from $u$ to $v$ with the smallest number of registers, denoted by $W(u \rightsquigarrow v)$.
We define the delay of this path as $D(u \rightsquigarrow v) = \sum_{k \in u \rightsquigarrow v} \delta(k)$ ($\delta$ is discussed in \Cref{subsubsec:xgboost-timing-predictor}).
Given the target delay $T$, PipeRTL adds a timing constraint when a minimal-register path becomes longer than the target and therefore must retain at least one register after relocation.
The resulting constraint is shown in \Cref{eq:objective_constrain}.
\begin{equation}
    \label{eq:objective_constrain}
    W(u \rightsquigarrow v) + \Delta s(v) - \Delta s(u) > 0.
\end{equation}
We define the node pairs that require this condition as the timing constraint set $\mathcal{C}$.

The overall pipeline optimization problem is shown in \Cref{eq:pipeline_optimization}. The objective minimizes total register data capacity, while the constraints enforce legality and timing requirements.
\begin{figure}[h!]
\begin{equation}
    \label{eq:pipeline_optimization}
        \begin{aligned}
            {\min}~     & \displaystyle\sum\limits_{v \in V} \Delta s(v) \cdot (\sum_{s \prec v}\beta(e_{s, v}) - \sum_{s \succ v}\beta(e_{v, s})) \\
            \text{s.t.}~&\quad {w(e_{u, v}) + \Delta s(v) - \Delta s(u) \ge 0,  \forall e_{u, v} \in E}. \\
            &\quad {W(u \rightsquigarrow v) + \Delta s(v) - \Delta s(u) > 0, \forall u, v \in \mathcal{C}}.\\
        \end{aligned}
\end{equation}
\end{figure}

\Cref{eq:pipeline_optimization} can be transformed into a min-cost flow problem and solved in polynomial time. The corresponding dual form is shown in \Cref{tbl:dual}. This formulation is important because it turns the relocation problem into a global optimization over the IR rather than a sequence of local greedy moves.

\begin{table}[tb!]
    \centering
    \caption{Dual form of the min-cost-flow formulation derived from \Cref{eq:pipeline_optimization}.}
    \label{tbl:dual}
    \resizebox{.55\linewidth}{!}
    {
    \begin{tabular}{|c|c|}
    \hline
    \textbf{Primal Program} & \textbf{Dual Program} \\ \hline \hline
        \tabincell{c}{
            minimize total register cost \\ 
            $\text{min} \displaystyle\sum_{v \in V}\Delta s(v) \cdot w[v]$
        }    
        &   
        \tabincell{c}{
        minimize flow cost \\
        $\text{min} \displaystyle\sum_{e \in E}c[e] \cdot f(e)$}            
        \\ \hline
        \tabincell{c}{
            node weight $w[v]$ \\
            $\displaystyle\sum_{s \prec v}\beta(e_{s, v}) - \displaystyle\sum_{s \succ v}\beta(e_{v, s})$
        }
        & 
        \tabincell{c}{
            node supply \texttt{supply[$v$]}\\
            $\displaystyle\sum_{s \prec v}\beta(e_{s, v}) - \displaystyle\sum_{s \succ v}\beta(e_{v, s})$
        }             
        \\ \hline
        \tabincell{c}{
            \\
            edge weight constraint(\Cref{eq:weight}) \\
            $\Delta s(v) - \Delta s(u) \ge -w(e_{u, v})$ \\
            \\
            timing constraint(\Cref{eq:objective_constrain}) \\
            $\Delta s(v) - \Delta s(u) > -W(u \rightsquigarrow v)$
                
            
        }
        &
        \tabincell{c}{
            cost $c$ for $E$\\
            $c[e_{u, v}] = -w(u, v)$\\
            cost $c$ for $\mathcal{C}$\\
            $c[e^{\mathcal{C}}_{u, v}] = -W(u \rightsquigarrow v) + 1$\\
            flow capacity \\
            $0 \le f(e) \le \infty$
        }
        \\ \hline
    \end{tabular}
    }
\end{table}

As shown in \Cref{fig:ir}(\circled{4}), the solution is materialized in the IR as a \texttt{pipe.potential} operation on each affected combinational node.

\subsection{IR Lowering} 
\label{subsec:ir-lowering}

The optimized stage updates are materialized by lowering \texttt{pipe.potential} back into standard sequential structure. According to \Cref{eq:weight}, PipeRTL recomputes the register count on each def-use edge after optimization, removes the original \texttt{pipe.delay} and \texttt{pipe.bubble} operations, and reconstructs the updated sequential structure from the optimized edge weights.

As shown in \Cref{fig:ir}(\circled{5}), if $w'(e_{u, v}) \neq 0$, the corresponding \texttt{pipe.potential} is lowered to \texttt{pipe.delay}; otherwise it is lowered to \texttt{pipe.bubble}. From \Cref{fig:ir}(\circled{5}) to \Cref{fig:ir}(\circled{6}), \texttt{pipe.delay} is further lowered to the standard \texttt{seq} dialect as shift-register structure, while \texttt{pipe.bubble} is eliminated from the def-use chain and becomes a wire connection in the generated RTL.
The optimized design corresponds to \Cref{lst:t1}, which performs timing-constrained register relocation while reducing the total register data capacity from 24 bits to 16 bits.
The optimization is expressed, solved, and materialized within the IR transformation flow, after which the design continues through standard CIRCT lowering for SystemVerilog generation and downstream backend implementation.

As a result, PipeRTL fits naturally into existing compiler pipelines: it operates at the IR level where global transformations are tractable, uses a custom dialect to represent the sequential structure needed for legality and optimization, and lowers back into standard IR forms for compatibility with existing passes and backends.

\section{Evaluation}
\label{sec:evaluation}

\begin{table*}[tb]
    \small
    \centering
    \caption{Benchmark description of PipeRTL.}
    \label{tbl:benchmark}
    \resizebox{\linewidth}{!}
    {
    \renewcommand{\arraystretch}{1.06}
    {
    \begin{tabular}{|l|l|l|l|l|}
      \hline
      \textbf{Source}            & \textbf{Benchmark} & \textbf{Description}                                            & \textbf{Category}   & \textbf{Development}     \\ \hline \hline
      \multirow{2}{*}{BOOM \cite{Arch-2015TR-BOOM}}      & BranchPredictor    & The branch predictor of BOOM                                   & Module              & Human-written \\
                                                         & Cache          & Non-blocking data cache with a three-stage pipeline                                        & Memory              & Human-written                     \\ \hline
      \multirow{4}{*}{Xiangshan\cite{micro2022xiangshan}} & StoreQueue         & Store queue between cache and main memory                       & FIFO                & Human-written             \\
                                 & CSR                & Control and status register module of Xiangshan CPU             & Module            & Human-written                     \\
                                 & ICache             & Instruction cache module of Xiangshan CPU                  & Memory              & Human-written         \\ 
                                 & TLB                & The first-level TLB with a page table walker and LRU replacement                                  & Module              & Human-written                     \\ \cline{1-5} 
      \multirow{2}{*}{TensorLib\cite{jia2021tensorlib}}                 & GEMM               & The PE array for general matrix multiply                        & Systolic Array      &          Automated Generation               \\
                                                                        & Conv2D              & The PE array for 2D convolution                       & Systolic Array      &        Automated Generation                  \\ \hline
      Gemmini \cite{Arch-2021DAC-Gemmini}                   & Gemmini            & Matrix-multiply core of Gemmini for $C = A * B + D $   & SoC & Human-written          \\ \hline
      WinoGen\cite{li2024winogen}                   & WinoGen             & Winograd convolution IP generator for DNN acceleration & IP Generator        &                Human-written          \\ \hline
      \end{tabular}
    }
    }
\end{table*}

\begin{table}[!tb]
  \centering
  \caption{Benchmark statistics.}
  \label{tbl:benchmark_statistic}
  \resizebox{.5\linewidth}{!}{ 
  \begin{tabular}{|c|c|c|c|c|c|}
      \hline
      \textbf{Benchmark} & \textbf{IR Nodes} & \textbf{IR Edges} & \textbf{\#Reg}     & \textbf{\#Capacity}      &\textbf{Clock ($ps$)}\\ \hline \hline
      BranchPredictor    & 3096              & 8022              & 407                & 3392                 &   400   \\
      Cache              & 10426              & 27760              & 2980                & 15737                 &   650   \\
      StoreQueue         & 13825             & 31850             & 2507               & 11022                &650      \\
      CSR             & 4292	             & 9328	             & 415	              & 8161                &450     \\
      ICache     & 3322	             & 6878	             & 386	              &6662	                 &650    \\
      TLB                & 1606            & 3718              & 256                & 1156                 &600    \\
      GEMM               & 5015              & 11704             & 1632               & 17262                 &650     \\
      Conv2D             &7599	             &17820	             &1969	              &13567                  &750     \\
      Gemmini            & 35631             & 65905             & 4019               & 40330                 &800     \\
      WinoGen            & 2030              & 4635	             &271	                & 2826                  &500     \\ \hline
    \end{tabular}
  }
\end{table}

\subsection{Experimental Settings}
\label{subsec:experimental-settings}

We use \texttt{firtool} from CIRCT as the frontend baseline and evaluate the generated RTL using Synopsys Design Compiler (DC) \cite{Misc-dc}.
The role of DC in this section is not to replace compiler optimization, but to provide a consistent downstream implementation environment for measuring the quality of compiler-generated pipeline structures.
All timing, power, and area numbers therefore correspond to the same backend implementation flow, enabling a fair comparison between the original RTL and PipeRTL-transformed RTL.
PipeRTL itself is implemented with approximately $4$K lines of C++ code and all experiments are conducted on Ubuntu 20.04 using an Intel i7-10750H CPU at 2.60\,GHz with 64\,GB memory.

\minisection{Benchmark Description}
To demonstrate the scalability of PipeRTL, we select critical modules as benchmarks from open-source Chisel designs, including BOOM \cite{Arch-2015TR-BOOM}, Gemmini\cite{Arch-2021DAC-Gemmini}, Xiangshan \cite{micro2022xiangshan}, TensorLib \cite{jia2021tensorlib}, and WinoGen\cite{li2024winogen}.
\Cref{tbl:benchmark} shows the detailed benchmark description.
These designs span domains such as out-of-order microprocessors, customized systems-on-a-chip (SoCs), and deep learning accelerators.
The circuit implementations in our benchmarks are varied, including systolic arrays, circular FIFO queues with associative searches, and multi-ported registers.
All external modules in the benchmark designs are SRAMs, whose internal timing is not optimized by PipeRTL.
\Cref{tbl:benchmark_statistic} presents the IR statistics, where ``\texttt{IR Nodes}'' denotes the total number of CIRCT core dialect operations and ``\texttt{IR Edges}'' denotes the use-define-chain relations.
It also reports the total number of registers (``\texttt{\#Reg}'') and the total register data capacity (``\texttt{\#Capacity}'').
The target clock periods shown in the last column of \Cref{tbl:benchmark_statistic} are obtained by frequency sweeping under the same DC flow and ASAP7 PDK used in the rest of the evaluation.
For each benchmark, we scan the clock period and select a near-boundary feasible timing target for the backend flow.
The same target clock is then used for both the original RTL and the PipeRTL-transformed RTL to ensure a fair downstream comparison; as a result, a small number of original no-retiming runs can slightly exceed the target and are reported as such.

\minisection{Design Compiler Settings}
PipeRTL is evaluated with DC through the \textit{dc-tcl} interface.
This evaluation employs the \SI{7}{\nm}~process design kit ASAP7 \cite{EDA-2016clarkasap7}, equipped with a composite current source (CCS) timing model to characterize transistor delays accurately.
We use the command ``\texttt{compile\_ultra}'' to perform logic synthesis for each design.
We report Design Compiler dynamic power estimates using its default constant activity factor of $0.1$ for all nets.
Moreover, to enable retiming in DC, we use the ``\texttt{-timing -retime}'' option for timing-constrained retiming and ``\texttt{set\_optimize\_registers}'' to allow register relocation.

\begin{table}[tb!]
  \centering
  \caption{Performance analysis of timing prediction.}
  \label{tab:timing_prediction}
  \renewcommand{\arraystretch}{1.26}
  \resizebox{.55\linewidth}{!}{ 
  \begin{tabular}{|l|rrr|rrr|}
    \hline
    \multicolumn{1}{|c|}{\multirow{2}{*}{\textbf{Model}}} & \multicolumn{3}{c|}{\textbf{Operation Regression}}                                                        & \multicolumn{3}{c|}{\textbf{Path Regression}}                                                                \\ \cline{2-7} 
    \multicolumn{1}{|c|}{}                                & \multicolumn{1}{c|}{$\textbf{RMSE}$} & \multicolumn{1}{c|}{$\textbf{MAE}$} & \multicolumn{1}{c|}{$\textbf{R}^\textbf{2}$} & \multicolumn{1}{c|}{$\textbf{Kendall}$} & \multicolumn{1}{c|}{$\textbf{MAE}$} & \multicolumn{1}{c|}{$\textbf{R}^\textbf{2}$} \\ \hline \hline
    MLP                                                   & 78.081                             & 19.144                            & 0.995                            & 0.874                                 & 1027.483                          & 0.761                            \\
    Decision Tree                                        & 54.796                             & 14.209                            & 0.997                            & 0.881                                 & \textcolor{vgreen}{\textbf{735.953}}                           & 0.879                            \\
    SVR                                                   & 509.437                            & 115.399                           & 0.818                            & 0.497                                 & 2212.613                          & 0.429                            \\
    KNN                                                   & 571.181                            & 128.099                           & 0.604                            & 0.829                                 & 961.076                           & 0.811                            \\
    Gradient Boost                                        & 173.871                            & 43.048                            & 0.969                            & 0.865                                 & 1003.061                          & 0.744                            \\
    \textbf{XGBoost}                                               & \textcolor{vgreen}{\textbf{42.041}}                             & \textcolor{vgreen}{\textbf{9.678}}                             & \textcolor{vgreen}{\textbf{0.997}}                            & \textcolor{vgreen}{\textbf{0.887}}                                 & 736.158                           & \textcolor{vgreen}{\textbf{0.881}}                            \\ \hline
    \end{tabular}
  }
\end{table}

\begin{figure*}[tb!]
  \centering
  \includegraphics[width=.88\linewidth]{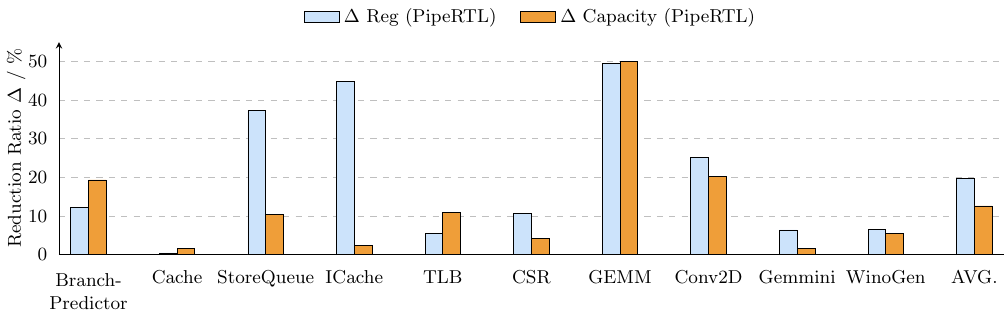}
  \caption{Register and register data capacity reduction of \textit{PipeRTL}. The average reduction across all benchmarks is \textbf{19.8\%} for register count and \textbf{12.6\%} for register data capacity.}
  \label{fig:bar}
\end{figure*}

\minisection{PPA Analysis Metrics}
\label{minisec:backend-ppa-analysis}

We use the following four metrics in our experiments.

\begin{description}
  \item [\textit{Critical Path Delay ($ps$)}] refers to the longest path delay reported by the backend implementation flow.
  \item [\textit{Total Dynamic Power ($mW$)}] represents the switching-power estimate reported by DC under the default activity model.
  \item [\textit{Cell Leakage Power ($\mu W$)}] measures static cell leakage reported by the backend library model.
  \item [\textit{Total Cell Area (${\mu m}^2$)}] refers to the physical area occupied by the cells on the chip die.
\end{description}

We use the geometric mean to aggregate results because it reduces skew from extreme values across different design sizes.

\begin{table*}[tb]
    \small
    \centering
  \caption{Visualization of the critical paths for the original and PipeRTL-optimized designs for \textit{GEMM}, \textit{Conv2D}, and \textit{WinoGen}.}
  \label{tbl:path_visualization}
  \resizebox{0.8\linewidth}{!}{
  \begin{tabular}{llll}
      \hline
      \multicolumn{1}{|c|}{\textbf{Benchmark}} & \multicolumn{1}{c|}{\textbf{Original Path} \tnote{1}}                       & \multicolumn{1}{c|}{\textbf{PipeRTL}}              & \multicolumn{1}{c|}{\textbf{PipeRTL (w.o $\mathcal{C}$)}}                                                              \\ \hline \hline
      \multicolumn{1}{|l|}{GEMM}      & \multicolumn{1}{l|}{$ \times (16b) \rightarrow +(16b) \rightarrow M(16b)$} & \multicolumn{1}{l|}{$ \times(16b) \rightarrow +(16b)$} & \multicolumn{1}{l|}{\begin{tabular}[c]{@{}l@{}}$ \times (16b) \rightarrow +(16b) \rightarrow M(16b)$\end{tabular}} \\ \hline
      \multicolumn{1}{|l|}{Conv2D}          & \multicolumn{1}{l|}{$\times(7b) \rightarrow +(7b)$}                                    & \multicolumn{1}{l|}{$ M(16b) \xrightarrow{\text{e}} +(9b)$}                         & \multicolumn{1}{l|}{$ \times(7b) \rightarrow 7 \odot +(7b)  $}                                                                          \\ \hline
      \multicolumn{1}{|l|}{WinoGen}          & \multicolumn{1}{l|}{$ +(4b) \rightarrow 2 \odot M(4b) $}                                    & \multicolumn{1}{l|}{$ +(4b) \rightarrow 2 \odot M(4b) $}                         & \multicolumn{1}{l|}{$ +(2b) \xrightarrow{\text{c}} 6 \odot M(8b) \rightarrow 2 \odot -(8b)$}                                                                          \\ \hline
  \end{tabular}
  }
  \begin{tablenotes}
      \small
      \item[1] $b$ bitwidth; $\times$ multiplier; $+$/$-$ adder/subtractor; $M$ multiplexer; $n \odot$ $n$ identical cascaded operations; $\rightarrow$ wire; \\ $\xrightarrow{\text{c/e}}$ bit concatenation/extraction.
  \end{tablenotes}
\end{table*}

\begin{table*}[!tbh]
  \centering
  \caption{{Downstream implementation quality of the original RTL and PipeRTL-transformed RTL, with and without backend retiming.}}
  \label{tab:result}
  \renewcommand{\arraystretch}{1.25}
  \resizebox{\linewidth}{!}{
      \begin{threeparttable}
        \begin{tabular}{|c|cccc|cccc|cccc|cccc|}
          \hline
          \multirow{3}{*}{\textbf{Benchmark}} & \multicolumn{4}{c|}{\textbf{Critical Path Delay ($ps$)}} & \multicolumn{4}{c|}{\textbf{Total Dynamic Power ($mW$)}} &  \multicolumn{4}{c|}{\textbf{Cell Leakage Power ($\mu W$)}} & \multicolumn{4}{c|}{\textbf{Total Cell Area (${\mu m}^2$)}} \\ \cline{2-17}
          & \multicolumn{2}{c|}{\textbf{w.o Retiming}} & \multicolumn{2}{c|}{\textbf{w. Retiming}} & \multicolumn{2}{c|}{\textbf{w.o Retiming}} & \multicolumn{2}{c|}{\textbf{w. Retiming}} & \multicolumn{2}{c|}{\textbf{w.o Retiming}} & \multicolumn{2}{c|}{\textbf{w. Retiming}} & \multicolumn{2}{c|}{\textbf{w.o Retiming}} & \multicolumn{2}{c|}{\textbf{w. Retiming}} \\ \cline{2-17}
          & Original & PipeRTL & Original & PipeRTL & Original & PipeRTL & Original & PipeRTL & Original & PipeRTL & Original & PipeRTL & Original & PipeRTL & Original & PipeRTL \\ \hline \hline
          BranchPredictor & 428.34 & 381.50 & 381.24 & 366.93 & 9.371 & 8.634 & 9.523 & 9.323 & 1.665 & 1.233 & 1.538 & 1.419 & 2037.322 & 1706.924 & 2126.289 & 1957.219 \\
          Cache & 630.78 & 630.85 & 630.65 & 628.94 & 267.751 & 267.623 & 267.828 & 267.661 & 51.448 & 45.023 & 45.764 & 45.850 & 68730.731 & 63644.556 & 63776.389 & 63721.203 \\
          StoreQueue & 641.53 & 642.29 & 648.84 & 629.20 & 19.33 & 16.705 & 15.475 & 15.243 & 4.996 & 4.607 & 4.279 & 4.215 & 6896.471 & 6274.634 & 5897.362 & 5800.230 \\
          CSR & 471.54 & 442.23 & 444.87 & 434.61 & 29.829 & 25.455 & 30.579 & 27.245 & 4.993 & 4.145 & 4.968 & 4.264 & 6779.423 & 5889.693 & 6903.295 & 5940.329 \\
          ICache & 637.61 & 623.06 & 632.90 & 632.99 & 12.729 & 12.259 & 11.374 & 11.321 & 2.674 & 2.393 & 2.647 & 2.802 & 3587.832 & 3235.739 & 3295.576 & 3287.702 \\
          TLB & 578.22 & 576.28 & 574.38 & 565.33 & 2.213 & 2.210 & 2.439 & 2.213 & 0.420 & 0.419 & 0.472 & 0.444 & 623.062 & 621.502 & 691.646 & 653.753 \\
          GEMM & 641.14 & 632.52 & 628.36 & 629.16 & 31.039 & 26.526 & 33.010 & 28.276 & 7.324 & 6.793 & 7.700 & 6.930 & 9643.912 & 8933.807 & 10388.279 & 9177.570 \\
          Conv2D & 731.96 & 731.37 & 741.93 & 715.49 & 57.936 & 55.631 & 58.181 & 55.873 & 12.052 & 11.799 & 12.726 & 12.238 & 17236.286 & 16774.304 & 17959.235 & 17378.879 \\
          Gemmini & 784.89 & 779.29 & 779.18 & 779.10 & 62.613 & 62.815 & 67.499 & 54.403 & 33.666 & 34.099 & 29.207 & 26.023 & 47314.389 & 46978.932 & 40704.910 & 37125.885 \\
          WinoGen & 479.14 & 476.69 & 480.88 & 449.16 & 5.923 & 5.572 & 5.950 & 5.822 & 1.058 & 1.008 & 1.104 & 1.084 & 1446.015 & 1405.045 & 1514.687 & 1513.491 \\ \hline \hline
          GEOMEAN & 592.35 & 578.78 & 581.32 & 569.00 & 22.05 & 20.60 & 21.94 & 20.48 & 4.87 & 4.43 & 4.75 & 4.51 & 6639.35 & 6169.28 & 6547.29 & 6191.40 \\
          Ratio Avg. & 100.00\% & \textcolor{darkgreen}{\textbf{97.71\%}} & 98.14\% & \textcolor{darkgreen}{\textbf{96.06\%}} & 100.00\% & \textcolor{darkgreen}{\textbf{93.41\%}} & 99.49\% & \textcolor{darkgreen}{\textbf{92.88\%}} & 100.00\% & \textcolor{darkgreen}{\textbf{90.97\%}} & 97.58\% & \textcolor{darkgreen}{\textbf{92.56\%}} & 100.00\% & \textcolor{darkgreen}{\textbf{92.92\%}} & 98.61\% & \textcolor{darkgreen}{\textbf{93.25\%}} \\ \hline
        \end{tabular}
      \end{threeparttable}
  }
\end{table*}

\subsection{Timing Model Analysis}
\label{subsec:timing-model-analysis}

We first evaluate whether the learned timing model is accurate enough to serve as the cost model for IR-level pipeline optimization.
This experiment is important because PipeRTL relies on downstream timing estimation to guide register relocation before lowering.
If the timing model is inaccurate, the global optimization may favor structurally legal but downstream-unprofitable solutions.

To evaluate the XGBoost-based timing prediction model, we compare it with several machine-learning algorithms, as shown in \Cref{tab:timing_prediction}.
We sample the \texttt{comb} dialect training dataset (500 samples) using Latin hypercube sampling.
First, we evaluate the regression performance of these algorithms on single combinational operations (1000 samples) of the \texttt{comb} dialect. 
Then, we randomly generate a combinational path dataset (500 samples) to evaluate the regression performance of these algorithms on path delay estimation.
Notably, the labels for the training and test datasets in these evaluations are generated by DC with the same settings and PDK as described in \Cref{subsec:experimental-settings}. 
Because these benchmarks are modeled as pure combinational logic, we set signal delays from all inputs to outputs to their theoretical minimum of zero, with no clock cycles.

The results are shown in \Cref{tab:timing_prediction}. The evaluation metrics include \textit{Root Mean Square Error} (RMSE), \textit{Mean Absolute Error} (MAE), \textit{$R^2$ score}, and \textit{Kendall's Tau}.
For operation regression, XGBoost performs best in terms of RMSE, MAE, and $R^2$, indicating that it captures operation timing effectively in our setting.
For path regression, while the MAE of XGBoost is slightly higher than that of the best-performing Decision Tree, XGBoost achieves the best scores in \textit{Kendall's Tau} and $R^2$.
This result further demonstrates the predictive accuracy of XGBoost and supports using summed operation delays as an approximation of path delay.
Overall, XGBoost proves to be the most suitable model for timing prediction in PipeRTL.

\subsection{Pipeline Optimization Quality}
\label{subsec:generated-verilog-evaluation}

We next evaluate the transformed RTL produced by PipeRTL at the same target clock period as the original design.
The goal of this experiment is to determine whether IR-level pipeline optimization improves both structural properties and downstream implementation quality.
Unless otherwise stated, all results in this subsection compare the original RTL and PipeRTL under the target clocks listed in \Cref{tbl:benchmark_statistic}.

\minisection{Structural Effects}
\label{subsec:Register-And-Bits-Reduction}
\Cref{fig:bar} illustrates the reduction ratios of register count and register data capacity achieved by PipeRTL over the original design. 
PipeRTL achieves an average register-count reduction of $19.84\%$ and a data-capacity reduction of $12.61\%$, demonstrating its effectiveness in reducing register usage at the RTL stage.
For \textit{TLB}, \textit{BranchPredictor}, and \textit{StoreQueue}, PipeRTL reduces register count and data capacity by roughly $10\%$ to $20\%$, showing that the transformation is effective on general-purpose processor blocks.
Furthermore, \textit{GEMM} and \textit{Conv2D} show even larger reductions in both metrics.
This suggests that the optimization is especially effective on regular dataflow structures with replicated modules.

\begin{table}[!tb]
  \centering
    \caption{Original and optimized critical-path delays predicted by the XGBoost model for each benchmark (all values are in $ps$).}
  \label{tbl:optimized_critical_path}
  \renewcommand{\arraystretch}{1.10}
  {
  \resizebox{.58\linewidth}{!}
  { 
  \begin{tabular}{|c|c|c|c|}
      \hline
      \textbf{Benchmark} & \textbf{Original Delay} & \textbf{Target Delay $T$} & \textbf{Optimized Delay} \\ \hline \hline
      BranchPredictor     & 223.83                          & 209.21                      & 209.19                            \\
      Cache    & 444.63	                        & 431.69	                    & 431.66	                          \\
      StoreQueue          & 371.36                          & 369.73                      & 369.69                            \\
      ICache              & 184.68                          & 171.74                      & 171.71                            \\
      CSR      & 549.43	                         & 536.49	                    & 536.46	                          \\
      TLB      & 158.43	                        & 140.56	                   & 136.89	                            \\
      GEMM                & 402.01                          & 389.05                      & 389.04                            \\
      Conv2D              & 227.38                            & 195.77                      & 195.75                              \\
      Gemmini             & 558.69                          & 532.77                      & 532.75                            \\
      WinoGen             & 257.13                          & 257.13                      & 257.13                            \\ \hline
  \end{tabular}
  }
  \begin{tablenotes}
     \small
    \item[1] The term ``Delay'' in this table refers to the predicted critical path delay.
\end{tablenotes}
  }
\end{table}

\minisection{Critical Path Analysis}
\label{minisec:critical-path-analysis}
To understand how the compiler transformation changes timing behavior before backend optimization, we inspect the predicted critical-path delay of the original and optimized IR solutions.
This analysis complements the downstream PPA numbers by showing that PipeRTL not only removes registers, but also reshapes the global timing structure seen by the backend.
We present the XGBoost predictions of the original and optimized critical-path delay for each benchmark in \Cref{tbl:optimized_critical_path}.
The third column shows the target delay $T$ obtained from the target-delay search described in \Cref{subsec:pipeline-optimization-framework}.
The last column displays the optimized critical path delay prediction for PipeRTL.
As shown in \Cref{tbl:optimized_critical_path}, the predicted optimized critical path delay is always equal to or smaller than the predicted original critical path delay.
This indicates that PipeRTL can either preserve the original critical path or further optimize it.

\Cref{tbl:path_visualization} shows the detailed critical paths for \textit{GEMM}, \textit{Conv2D}, and \textit{WinoGen} in the original and PipeRTL-optimized designs.
For \textit{GEMM}, PipeRTL effectively shortens the original critical path by realigning the pipeline of the cascaded multiplexer tree.
For \textit{Conv2D}, the critical path of the PipeRTL-optimized design shifts to a different path relative to the original design. 
This shows that PipeRTL does not merely preserve local path balance; it can change which path becomes critical and therefore acts as a global pipeline optimization rather than a local heuristic.
For \textit{WinoGen}, PipeRTL preserves the original critical path during optimization.

\begin{table*}[tb!]
  \centering
  \caption{Ablation study of PipeRTL performance with and without timing constraints (\textit{w.} retiming).}
  \label{tab:res_hint}
  \renewcommand{\arraystretch}{1.26}
  \resizebox{\linewidth}{!}{
      \begin{threeparttable}
        \begin{tabular}{|c|ccc|ccc|ccc|ccc|}
          \hline
          \multirow{2}{*}{\textbf{Benchmark}} & \multicolumn{3}{c|}{\textbf{Critical Path Delay ($ps$)}} & \multicolumn{3}{c|}{\textbf{Total Dynamic Power ($mW$)}} &  \multicolumn{3}{c|}{\textbf{Cell Leakage Power ($\mu W$)}} & \multicolumn{3}{c|}{\textbf{Total Cell Area (${\mu m}^2$)}} \\ \cline{2-13}
          & Origin & PipeRTL (w.o $\mathcal{C}$) & PipeRTL & Origin & PipeRTL (w.o $\mathcal{C}$) & PipeRTL & Origin & PipeRTL (w.o $\mathcal{C}$) & PipeRTL & Origin & PipeRTL (w.o $\mathcal{C}$) & PipeRTL \\ \hline \hline
          StoreQueue & 648.84 & 628.03 & 629.20 & 15.475 & 16.041 & 15.243 & 4.279 & 4.359 & 4.215 & 5897.362 & 6027.882 & 5800.230 \\
          Conv2D & 741.93 & 729.38 & 715.49 & 58.181 & 55.316 & 55.873 & 12.726 & 12.284 & 12.238 & 17959.235 & 17352.197 & 17378.879 \\
          GEMM & 628.36 & 628.91 & 629.16 & 33.010 & 31.931 & 28.276 & 7.700 & 7.504 & 6.930 & 10388.279 & 10118.986 & 9177.570 \\
          Gemmini & 779.18 & 1472.62 & 779.10 & 67.499 & 51.105 & 54.403 & 29.207 & 25.759 & 26.023 & 40704.910 & 35308.371 & 37125.885 \\
          WinoGen & 480.88 & 479.28 & 449.16 & 5.950 & 6.724 & 5.822 & 1.104 & 1.240 & 1.084 & 1514.687 & 1742.981 & 1513.491 \\
          TLB & 574.38 & 577.75 & 565.33 & 2.439 & 2.432 & 2.213 & 0.472 & 0.469 & 0.444 & 691.646 & 689.109 & 653.753 \\ \hline \hline
          GEOMEAN & 634.25 & 699.83 & 618.60 & 17.54 & 16.95 & 16.02 & 4.31 & 4.26 & 4.06 & 6005.61 & 5961.46 & 5691.29 \\
          Ratio Avg. & 100.00\% & \textcolor{red}{\textbf{110.34\%}} & \textcolor{darkgreen}{\textbf{97.53\%}} & 100.00\% & \textcolor{vgreen}{\textbf{96.62\%}} & \textcolor{darkgreen}{\textbf{91.32\%}} & 100.00\% & \textcolor{vgreen}{\textbf{99.03\%}} & \textcolor{darkgreen}{\textbf{94.27\%}} & 100.00\% & \textcolor{vgreen}{\textbf{99.26\%}} & \textcolor{darkgreen}{\textbf{94.77\%}} \\ \hline
        \end{tabular}
      \end{threeparttable}
  }
\end{table*}

\minisection{PPA Evaluation w.o. Retiming}
\label{minisec:evaluation-without-retime}
We first evaluate PipeRTL without enabling backend retiming in DC.
This experiment isolates the quality of the compiler transformation itself: any improvement here comes from IR-level pipeline optimization rather than from backend sequential optimization.
In other words, this setting answers whether the compiler pass alone can generate a better sequential structure for downstream synthesis.
The results are presented in the columns under ``\textbf{w.o Retiming}'' in \Cref{tab:result}.

Based on the above analysis of register and data capacity reduction, PipeRTL reduces both total cell area and total dynamic power on average.
Although PipeRTL does not explicitly target cell leakage power, it also reduces leakage on average relative to the original RTL.
This reduction is consistent with the decrease in register count and data capacity, which can expand the optimization space for combinational logic; such power and area gains are meaningful under the fixed backend flow, functionality, timing constraints, and technology mapping used in this study.

Since timing serves as a constraint rather than the sole optimization objective in PipeRTL, PipeRTL still achieves an average timing improvement of about $2.3\%$ even before backend retiming.
This result is important in a compiler setting: rather than merely shifting cost across metrics, PipeRTL produces a better starting point for downstream implementation by reducing area and power without sacrificing timing robustness in the evaluated flow.

In particular, the original \textit{BranchPredictor} design violates the target clock of $400~ps$, while the PipeRTL-optimized design meets timing with lower power and area.
As shown in \Cref{fig:bar}, \textit{Cache} and \textit{ICache} exhibit only modest reductions in register count and capacity, yet still obtain meaningful area and power improvements. This highlights the downstream impact of changing the sequential structure, not just the raw number of registers.

\begin{figure*}[tb!]
	\centering
	\includegraphics[width=\linewidth]{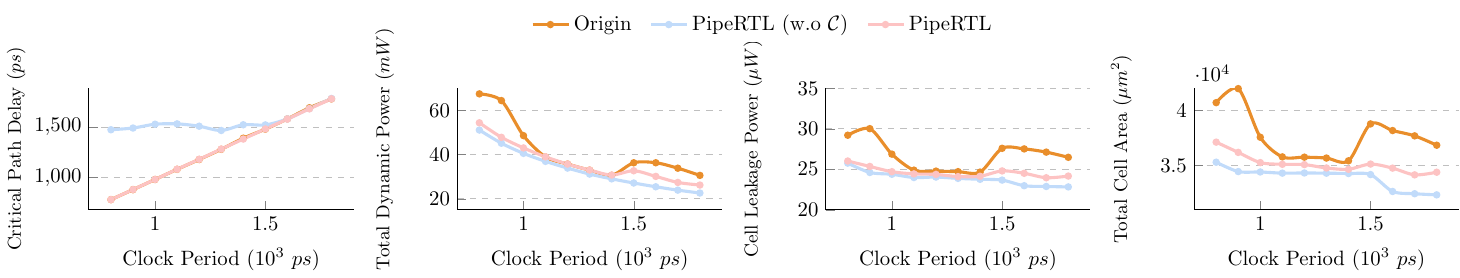}
	\caption{PPA evaluation of \textit{Gemmini} under a target-clock sweep.}
	\label{fig:gemmini_case_study}
\end{figure*}

\minisection{Co-optimization with Retiming}
\label{minisec:evaluation-with-retime}
We next evaluate the generated RTL with retiming enabled in DC to determine whether IR-level optimization can provide additional benefits beyond backend sequential optimization in our evaluated flow.
The results are presented in the columns under ``\textbf{w. Retiming}'' in \Cref{tab:result}.
Because logic synthesis has more accurate physical information, the PPA of both the original RTL and the PipeRTL-transformed RTL can be further improved after retiming.
First, as shown in \Cref{tab:result}, PipeRTL without retiming already outperforms the original design with retiming on several benchmarks, indicating that the compiler pass can improve the starting sequential structure in some cases.
Furthermore, compared with the original design with retiming, PipeRTL with retiming improves timing, area, and power on average across the evaluated benchmarks.
This finding is central to our argument: IR-level pipeline optimization does not compete with backend retiming for the same optimization budget.
Instead, it provides a better sequential starting point on top of which backend retiming can continue refining the design.
Overall, compared with the original RTL with retiming, PipeRTL with retiming achieves about a $2.1\%$ timing improvement together with reductions of $6.6\%$ in total dynamic power, $5.1\%$ in cell leakage power, and $5.4\%$ in total cell area in the evaluated backend flow.
This indicates that preserving pipeline semantics before lowering exposes optimization opportunities that remain valuable even when backend sequential optimization is enabled.

\begin{table}[!tb]
  \centering
  \caption{Predicted critical-path delay for PipeRTL (w.o $\mathcal{C}$) and the reduction in register count ($\Delta$Reg) and data capacity ($\Delta$Capacity) relative to PipeRTL.}
  \label{tab:delta_metrics}
  \renewcommand{\arraystretch}{1.02}
  {
  \resizebox{0.5\linewidth}{!}
  { 
  \begin{tabular}{|c|c|c|c|}
      \hline
      \textbf{Benchmark} & \textbf{Optimized Delay} & \textbf{$\Delta$Reg} & \textbf{$\Delta$Capacity} \\ \hline \hline
      StoreQueue & \textcolor{red}{448.79} & $+1.64\%$ & $+0.42\%$ \\
      Conv2D     & \textcolor{red}{642.76} & $+8.58\%$ & $+4.43\%$ \\
      GEMM       & 402.01 & 0.00\% & 0.00\% \\
      Gemmini    & \textcolor{red}{5,874.58} & $+14.26\%$ & $+4.47\%$ \\
      WinoGen    & \textcolor{red}{264.31} & 0.00\% & 0.00\% \\
      TLB        & \textcolor{red}{249.45} & $+1.56\%$ & $+0.61\%$ \\ \hline
  \end{tabular}
  }
  }
\end{table}

\subsection{Necessity of Timing Constraints}
\label{minisec:timing-constraint-hints}
We then study whether timing constraints are necessary in the IR optimization itself.
This ablation is important because PipeRTL does not optimize purely for register reduction; it also needs to preserve a legal and profitable timing-aware search space.
To evaluate this effect, we remove the timing constraint in \Cref{eq:objective_constrain} on several benchmarks and compare the resulting RTL with the full system under the same downstream retiming configuration.
We define PipeRTL with timing constraints as PipeRTL and without timing constraints as PipeRTL (w.o $\mathcal{C}$).

\Cref{tab:res_hint} shows the PPA results without timing constraints.
PipeRTL (w.o $\mathcal{C}$) can hurt timing, area, and power relative to both PipeRTL and the original design.
As shown in \Cref{tab:delta_metrics}, although PipeRTL (w.o $\mathcal{C}$) can achieve larger reductions in register count and data capacity because it has fewer constraints, the critical path delay may degrade substantially.
The last column of \Cref{tbl:path_visualization} visualizes the optimized critical paths without timing constraints.
For \textit{Conv2D}, PipeRTL (w.o $\mathcal{C}$) disrupts the pipeline structure of the adder tree, causing each level of the tree to cascade together, thereby extending the original critical path.
Furthermore, for \textit{WinoGen}, other paths in the circuit are disrupted and become a new critical path.
This ablation demonstrates that legality-preserving structure alone is insufficient: the compiler also needs a timing-aware objective to avoid transformations that are structurally attractive but downstream harmful.

Specifically, for \textit{StoreQueue} and \textit{WinoGen}, total cell area and power are even larger than in the original design, as shown in \Cref{tab:res_hint}.
One plausible reason is the use of heuristic algorithms in DC during logic synthesis.
Notably, some combinational optimization techniques, such as logic restructuring \cite{lou1999timing,yang2012timing,brenner2022timing} or sum-of-products balancing \cite{alan2011timing}, can lead to better timing adjustments at the expense of increased area and power consumption.
The timing results of these two designs show no violations, which is consistent with this interpretation.
Although the critical path of \textit{GEMM} can be preserved without timing constraints, as shown in \Cref{tbl:path_visualization}, PipeRTL further improves PPA by shortening the critical path, as shown in \Cref{tab:res_hint}.
Furthermore, even with retiming applied, the \textit{Gemmini} design fails to meet the timing requirement without timing constraints, because it significantly disrupts the critical path by removing the registers in the high-bitwidth multiplexer tree compared with the original design, as shown in \Cref{tbl:optimized_critical_path} and \Cref{tab:delta_metrics}.

The overall average improvement in PPA shown in \Cref{tab:res_hint} therefore supports the role of timing constraints as a necessary component of the compiler optimization objective, not merely an optional heuristic hint.

\begin{table}[!tb]
  \centering
  \caption{Post Place-and-Route (PnR) results of \textit{Gemmini} through Innovus with retiming enabled.}
  \label{tbl:pnr_gemmini}

  \renewcommand{\arraystretch}{1.13}
  {
  \resizebox{0.8\linewidth}{!}
  { 
  \begin{tabular}{|c|c|c|c|c|}
      \hline
      \textbf{Condition} & \textbf{Wire Length ($\mu m$)} & \textbf{\#Standard Cell} & \textbf{Area ($mm^{2}$)} & \textbf{Power ($mW$)} \\ \hline \hline
      w.o. PipeRTL         & \textcolor{darkgreen}{5606517}	& 492021	& 0.7479 	& 1.2452	                     \\ 
      w. PipeRTL       & 5975710 	& \textcolor{darkgreen}{474437}	& \textcolor{darkgreen}{0.7319}	& \textcolor{darkgreen}{1.0658}	                         \\ \hline 
  \end{tabular}
  }
  \begin{tablenotes}
      \small
      \item[1] The target clock is the same as in \Cref{tbl:benchmark_statistic}.\\
      The generated layout has no timing violations (both \textit{WNS} and \textit{TNS} are 0).
  \end{tablenotes}
  }
\end{table}

\subsection{Case Study: Stability Across Target Clocks}
\label{subsec:case-study}

\Cref{fig:gemmini_case_study} shows the PPA evaluation of \textit{Gemmini} under a target-clock sweep from $800~ps$ to $1800~ps$ in $100~ps$ steps with retiming enabled.
When the target clock is below $1500~ps$, PipeRTL (w.o $\mathcal{C}$) does not meet the timing requirement, consistent with the results in \Cref{tab:res_hint}. In contrast, PipeRTL always meets the requirement and achieves slightly better timing than the original RTL.
When the target clock is set to $900~ps$ and $1500~ps$, the power and area curves of the original design exhibit a bump.
This occurs because DC's timing constraints approach the critical-path threshold, causing the adopted heuristics to encounter a corner case and produce slightly inferior results.
In contrast, the PipeRTL-optimized design is more stable, indicating that the compiler-generated sequential structure is more robust across a wider range of backend timing targets.
When the target clock sweeps from $1100~ps$ to $1400~ps$, although the total dynamic power curve of PipeRTL is close to that of ``Origin'', total cell area and cell leakage power improve.
Overall, PipeRTL continues to improve several metrics across different target clock periods.

\Cref{tbl:pnr_gemmini} presents the post-place-and-route (PnR) results of \textit{Gemmini} using the commercial tool Innovus.
Compared with the original design, the PipeRTL-optimized design improves both chip area and power.
This demonstrates that the optimizations introduced by PipeRTL are preserved through the post-layout stage.

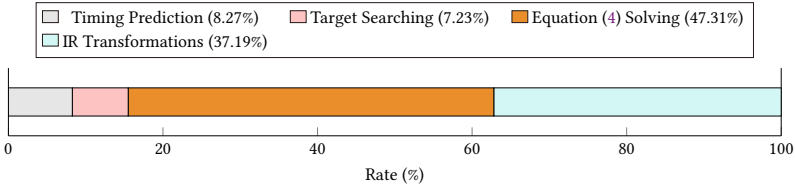
\begin{figure}[t]
  \centering
  \resizebox{0.78\linewidth}{!}
  { 
      \begin{tikzpicture}
    \pgfplotsset{
        width=1.1\linewidth,
        height=0.2\linewidth 
    }
    \definecolor{black}{RGB}{34,24,21}
    \definecolor{gray9}{gray}{0.9}
    \definecolor{YIOrRd4}{RGB}{252,163,60}
    \definecolor{Orange}{RGB}{233,142,43}
    \definecolor{Blue}{RGB}{193,219,250}
    \definecolor{Pink}{RGB}{253,195,195}
    \definecolor{Cyan}{RGB}{211,247,244}

    \begin{axis}[
        xbar stacked, xmin=0,
        bar width=0.5cm, 
        ytick={1},
        yticklabels={},
        ymin=0,
        ymax=2,
        xmin=0,
        xmax=100,
        xtick={0, 20, 40, 60, 80, 100}, 
        xticklabels={0, 20, 40, 60, 80, 100}, 
        xlabel={Rate (\%)},
        tick label style={font=\small},
        label style={font=\small},
        axis x line*=bottom, 
        xtick pos=bottom,
        legend style={at={(0.5,1.15)}, anchor=south, legend columns=3, font=\small, /tikz/every even column/.append style={column sep=0.33cm}},
    ]
    \addplot [fill=gray9] coordinates {(8.27,1)}; 
    \addplot [fill=Pink] coordinates {(7.23,1)}; 
    \addplot [fill=Orange] coordinates {(47.31,1)}; 
    \addplot [fill=Cyan] coordinates {(37.19,1)}; 
    
    \legend{Timing Prediction (8.27\%), Target Searching (7.23\%), \Cref{eq:pipeline_optimization} Solving (47.31\%), IR Transformations (37.19\%)}
    \end{axis} 
\end{tikzpicture}
  }
  \caption{Average compilation runtime breakdown of PipeRTL.}
  \label{fig:breakdown}
\end{figure}

\begin{table}[t]
  \centering
  \caption{Benchmark-wise runtime comparison of pipeline optimization in PipeRTL and retiming in DC.}
  \label{tbl:runtime_compare}

  \renewcommand{\arraystretch}{1.05}
  {
  \resizebox{0.5\linewidth}{!}
  { 
  \begin{tabular}{|c|c|c|c|}
      \hline
      \textbf{Benchmark} & \textbf{$|\mathcal{C}|$} & \textbf{PipeRTL ($s$)} & \textbf{Retiming ($s$)} \\ \hline \hline
      BranchPredictor & 143322 & 2.1 & 395 \\
      Cache & 2846269 & 62.2 & 3742 \\
      StoreQueue & 172674 & 11.7 & 888 \\
      ICache & 566239 & 7.0 & 422 \\
      TLB & 46932 & 3.3 & 167 \\
      CSR & 7319 & 0.7 & 1243 \\
      GEMM & 3751 & 0.7 & 875 \\
      Conv2D & 21954 & 0.4 & 1363 \\
      Gemmini & 1817383 & 16.1 & 3023 \\
      WinoGen & 28792 & 0.6 & 236 \\ \hline \hline
      GEOMEAN & 94866.2 & 3.0 & 792.6 \\
      Ratio Avg. & - & $1.0\times$ & $262.0\times$ \\ \hline
  \end{tabular}
  }
  }

\end{table}

\subsection{Runtime Overhead}
\label{subsec:runtime-analysis}
Finally, we analyze the runtime cost of PipeRTL as a compiler optimization pass.
Across the benchmarks in \Cref{tbl:runtime_compare}, PipeRTL runtime ranges from $0.4~s$ for \textit{Conv2D} to $62.2~s$ for \textit{Cache}.
\Cref{fig:breakdown} illustrates the geometric-mean runtime breakdown for benchmark compilation with PipeRTL.
We analyze the runtime of the core optimization phases in PipeRTL, including XGBoost-based timing prediction, target-delay search, solving the pipeline optimization objective, and IR transformation and lowering.
The breakdown shows that the most time-consuming phase is solving the pipeline optimization objective, since it requires solving min-cost flow problems that form the core of the global optimization process.

As shown in \Cref{tbl:runtime_compare}, pipeline optimization in PipeRTL is considerably faster than retiming optimization in DC, with an average speedup of about $262\times$, because it operates on a smaller problem at the RTL stage while still achieving measurable downstream improvements.
Notably, DC's retiming phase uses 16 threads, whereas PipeRTL's pipeline optimization runs single-threaded, further underscoring PipeRTL's computational efficiency.
The runtime of pipeline optimization in PipeRTL is dominated by min-cost flow solving, which is sensitive to the size of the constraint set $\mathcal{C}$.
This runtime profile suggests that PipeRTL is practical as an IR-level compiler pass, while still delivering meaningful downstream implementation gains and complementing backend sequential optimization.

\section{Related Work}
\label{sec:related-work}

\minisection{CIRCT-based Hardware Compilers}
CIRCT~\cite{Misc-CIRCT} is a hardware compiler infrastructure built on top of MLIR~\cite{Sys-2021CGO-MLIR} and has been increasingly adopted for hardware generation and optimization. 
Calyx~\cite{nigam2021compiler} is a hardware IR for accelerator generation, and LLHD~\cite{schuiki2020llhd} aims to unify the IR used in EDA.
HLS tools such as ScaleHLS~\cite{ye2022scalehls} and HIDA~\cite{ye2023hida} use core MLIR dialects from software compilation to perform loop-level transformations and optimizations before lowering to CIRCT for hardware generation.
Assassyn~\cite{weng2025assassyn} uses CIRCT to bridge gaps between the RTL generation, verification, and simulation stages of the hardware design process. 
Cement~\cite{xiao2024cement} defines event-based semantics for pipelined hardware design based on CIRCT.
PipeRTL is complementary to these efforts in that it adds an explicit IR-level pipeline optimization stage before lowering into less structured sequential forms. 
This broadens the optimization scope of CIRCT-based hardware compilers: instead of leaving pipeline organization entirely to frontend heuristics or backend retiming, these compilers can expose sequential structure in CIRCT and apply PipeRTL as a reusable middle-end pass.

\minisection{Retiming}
Retiming is a sequential optimization technique in logic synthesis, originally introduced in \cite{leiserson1991retiming}, with a formulation that parallels PipeRTL.
Unlike PipeRTL, which performs optimization directly at the IR level without gate-level synthesis, traditional retiming algorithms are rooted in logic-based analysis of synthesized netlists.
This analysis incorporates gate-level logic functions but often overlooks component bitwidth.
Building on this foundational concept, subsequent retiming work has aimed to improve efficiency and scalability.
\cite{shenoy2003efficient} and \cite{yu2011network} transformed the original problem into a flow-based problem, reducing its complexity.
\cite{hurst2007fast} further simplified the problem into a maximum-flow formulation, achieving a faster solution.
\cite{hurst2008scalable} proposed a tree-based constraint graph heuristic, which performs well in large circuit designs.
PipeRTL does not attempt to replace this line of work.
Instead, it brings a related optimization objective into the compiler middle-end, where operation identities, stage boundaries, and structural legality are still explicit.
This is why PipeRTL and backend retiming are complementary in our experiments rather than mutually exclusive alternatives.

\section{Conclusion}
\label{sec:conclusion}

This paper presented PipeRTL, a compiler-guided IR-level pipeline optimization framework for RTL generation.
Instead of leaving sequential optimization entirely to backend retiming after lowering, PipeRTL optimizes register placement while operation-level semantics and pipeline structure are still explicit.
To support this, we introduce a CIRCT dialect with explicit pipeline-legality semantics, use a learned downstream-aware timing predictor, and solve global register relocation as a min-cost flow problem under timing constraints.
Across the evaluated open-source processor and accelerator designs, PipeRTL improves downstream timing, power, and area on average, and provides additional gains when combined with backend retiming in our evaluated flow.
Overall, our results show that preserving optimization-relevant pipeline semantics in compiler IR enables effective cost-aware sequential optimization before lowering and provides a useful direction for future hardware compilers.

{
    \bibliographystyle{ACM-Reference-Format}
    \bibliography{refs/top,refs/ref}
}

\end{document}